\documentclass[12pt]{article}
\pdfoutput=1
\usepackage{amsmath,amssymb,amsfonts}
\usepackage{a4wide,epsfig,psfrag,scalefnt}
\usepackage[dvipsnames]{xcolor}
\usepackage{braket}
\usepackage{placeins}
\usepackage{breqn}
\usepackage{slashed}
\usepackage{enumitem}
\usepackage[numbers,sort&compress]{natbib}
\usepackage{caption}
\usepackage{subcaption}

\graphicspath{ {figs/} }

\allowdisplaybreaks

\begin{document}

\title{\vskip-3cm{\baselineskip14pt
    \begin{flushleft}
     \normalsize LTH 1371, P3H-24-034, TTP24-016, ZU-TH 27/24
    \end{flushleft}} \vskip1.5cm
  Three-loop corrections to Higgs boson pair production: reducible contribution
}
 
\author{
  Joshua Davies$^{a}$,
  Kay Sch\"onwald$^{b}$,
  Matthias Steinhauser$^{c}$,
  Marco Vitti$^{c,d}$,
  \\
  {\small\it (a) Department of Mathematical Sciences, University of Liverpool,
    Liverpool, L69 3BX, UK}
  \\
  {\small\it (b) Physik-Institut, Universit\"at Z\"urich, Winterthurerstrasse 190,}\\
  {\small\it 8057 Z\"urich, Switzerland}
  \\
  {\small\it (c) Institut f{\"u}r Theoretische Teilchenphysik,
     Karlsruhe Institute of Technology (KIT),}\\
  {\small\it Wolfgang-Gaede Stra\ss{}e 1, 76128 Karlsruhe, Germany}
  \\
  {\small\it (d) Institut f{\"u}r Astroteilchenphysik,
    Karlsruhe Institute of Technology (KIT),}\\
  {\small\it Hermann-von-Helmholtz-Platz 1, 76344 Eggenstein-Leopoldshafen, Germany, Germany}
}

\date{}

\maketitle

\thispagestyle{empty}

\begin{abstract}

  We compute three-loop corrections to the process $gg\to HH$
  originating from one-particle reducible diagrams. This requires the computation
  of two-loop corrections to the gluon-gluon-Higgs vertex with an
  off-shell gluon. We describe in detail our approach to obtain
  semi-analytic results for the vertex form factors and present results
  for the two form factors contributing to Higgs boson pair
  production.
  
\end{abstract}


\newpage


\section{Introduction and notation}

The simultaneous production of two Higgs bosons is the most promising process
from which we can obtain information about the Higgs boson's self coupling.  At
hadron colliders it is dominated by the gluon fusion production channel, which
at leading order (LO) receives contributions from ``triangle''- and
``box''-type diagrams.  The LO form factors and cross section were computed
more than 35 years ago~\cite{Glover:1987nx,Plehn:1996wb}. Next-to-leading
order (NLO) corrections with full dependence on the top quark mass are
available in numerical form from
Refs.~\cite{Borowka:2016ehy,Borowka:2016ypz,Baglio:2018lrj}. ``Semi-analytic
expressions'' which are essentially equivalent to the numerical
results, but more flexible in that mass values can be adjusted, have been computed in
Refs.~\cite{Bellafronte:2022jmo,Davies:2023vmj}.  Furthermore there are a
number of approximations which are valid in certain regions of phase space
(see, e.g.,
Refs.~\cite{Grigo:2013rya,Degrassi:2016vss,Davies:2018ood,Davies:2018qvx,Bonciani:2018omm,Grober:2017uho,Xu:2018eos,Wang:2020nnr}).

\begin{figure}[tb]
  \centering
  \includegraphics[width=.345\linewidth]{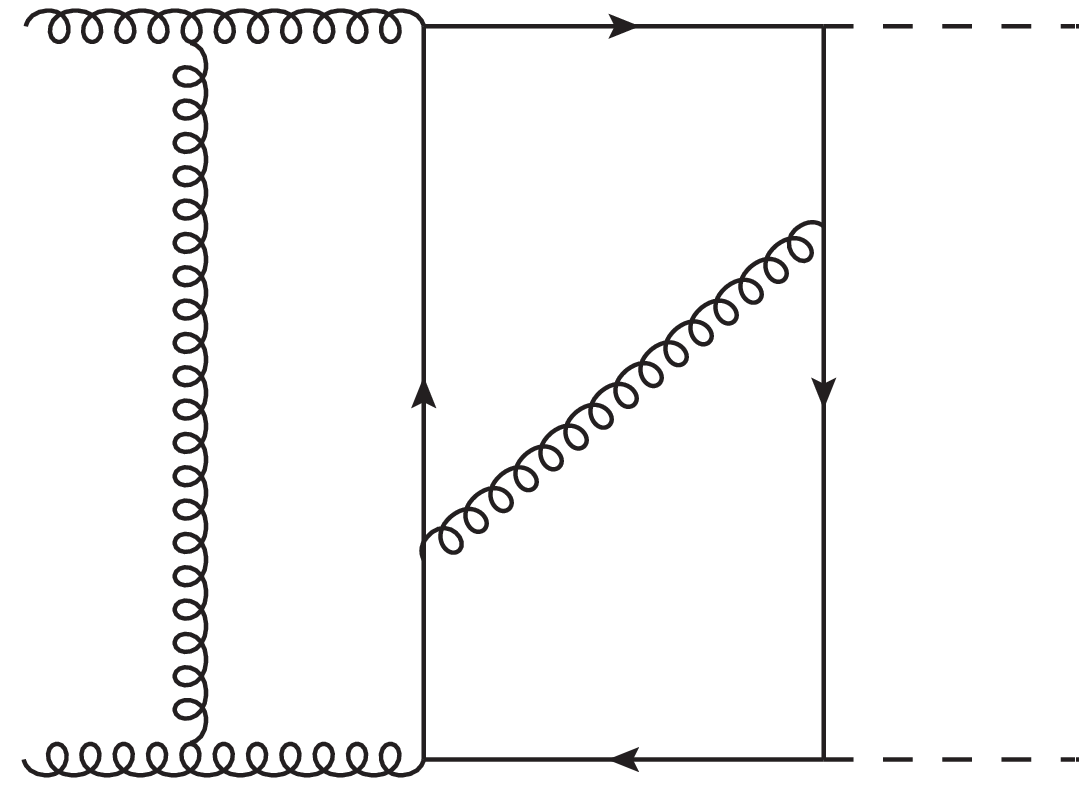}
  \includegraphics[width=.3\linewidth]{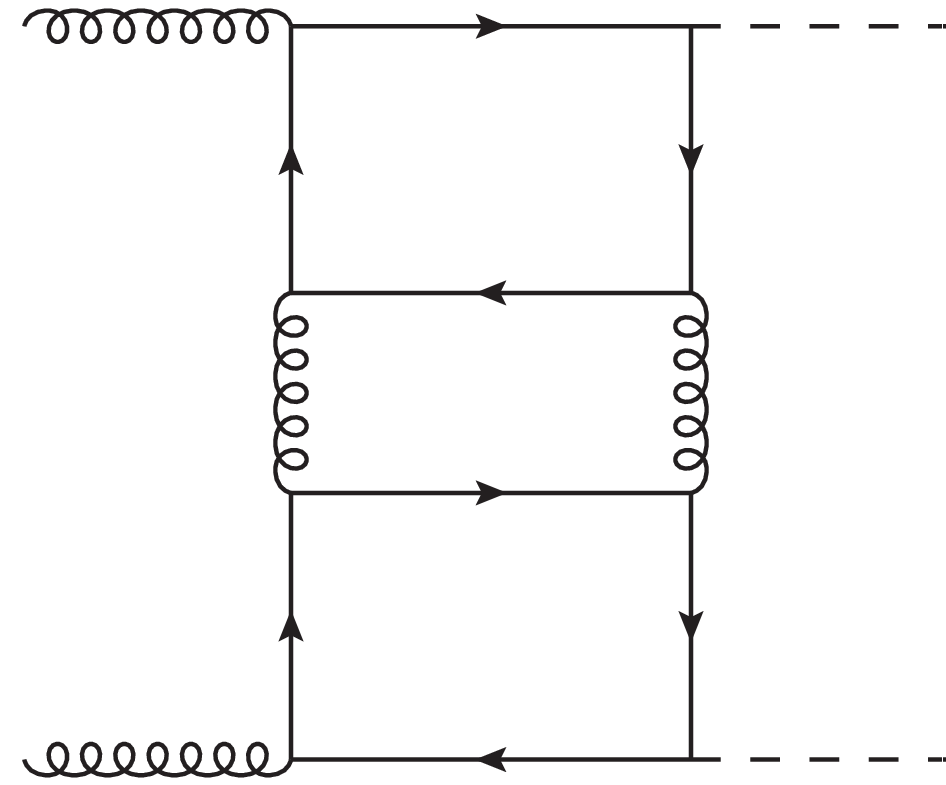}
  \includegraphics[width=.3\linewidth]{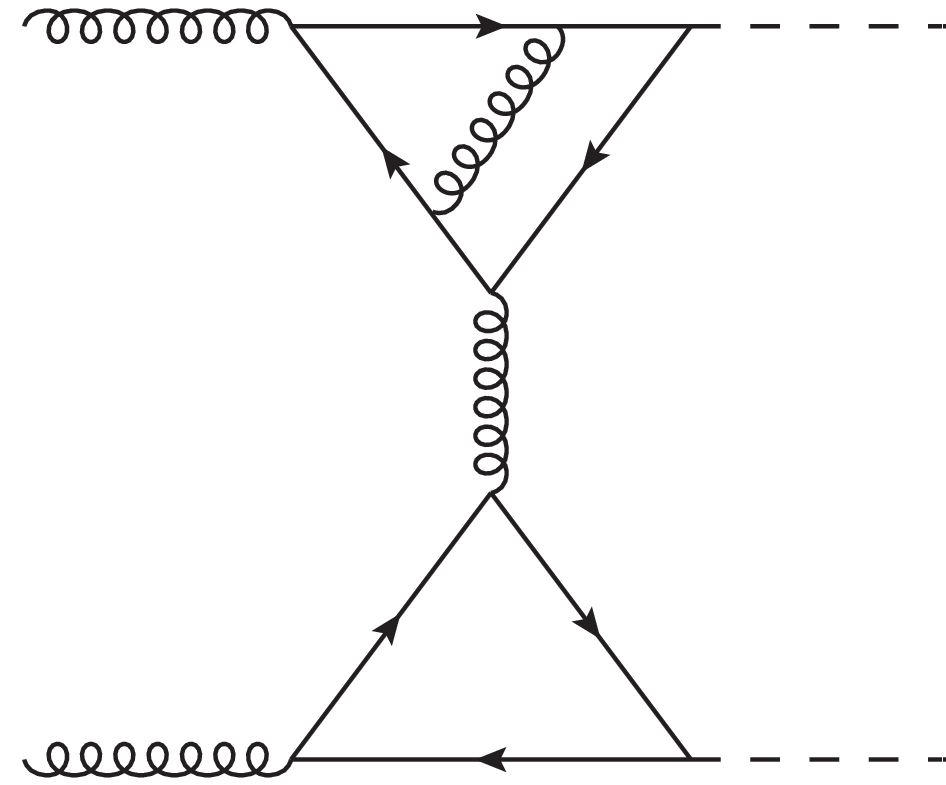}
  \\
  \mbox{}\hfill(a)\hfill\hspace*{3em}(b)\hfill\mbox{}\hspace*{6em}(c)\hfill\mbox{}
  \caption{Classification of the three-loop virtual corrections to $gg\to HH$. (a) shows ``class (i)'', where both Higgs bosons couple to the same top quark loop. (b) and (c) show ``class (ii)'', where each Higgs boson couples to a different top quark loop, in a one-particle irreducible and reducible way, respectively.}\label{fig::diag_gghh}
\end{figure}

In Refs.~\cite{Baglio:2020wgt,Bagnaschi:2023rbx} it has been pointed out that the renormalization
scheme dependence of the top quark mass induces a sizeable uncertainty on the
NLO Higgs boson pair cross section. This motivates a
next-to-next-to-leading order (NNLO) calculation of the Higgs boson pair
production cross section. The most challenging part in this context is the
three-loop virtual correction to $gg\to HH$. This can be divided into two classes: (i)
diagrams where both Higgs bosons couple to the same top quark loop (as shown
in Fig.~\ref{fig::diag_gghh}(a)), and (ii) diagrams where the Higgs bosons
couple to different top quark loops. For the latter class there are one-particle
irreducible and one-particle reducible contributions (as shown in
Figs.~\ref{fig::diag_gghh}(b) and (c) respectively).  The irreducible and
reducible contributions are, separately, neither finite nor gauge-parameter
independent, which has already been discussed in
Refs.~\cite{Davies:2019djw,Davies:2021kex}.

The large top quark mass limit of the complete set of diagrams has been
considered in
Refs.~\cite{Grigo:2015dia,Davies:2019djw}, where five expansion
terms in $1/m_t^2$ were computed. 
The light-fermion contribution to class (i) has
been computed in Ref.~\cite{Davies:2023obx} for $t=0$ and $m_H=0$, which
is a promising method to obtain an approximation of the unknown exact result.

\begin{figure}[b]
  \centering
  \includegraphics[width=.22\linewidth]{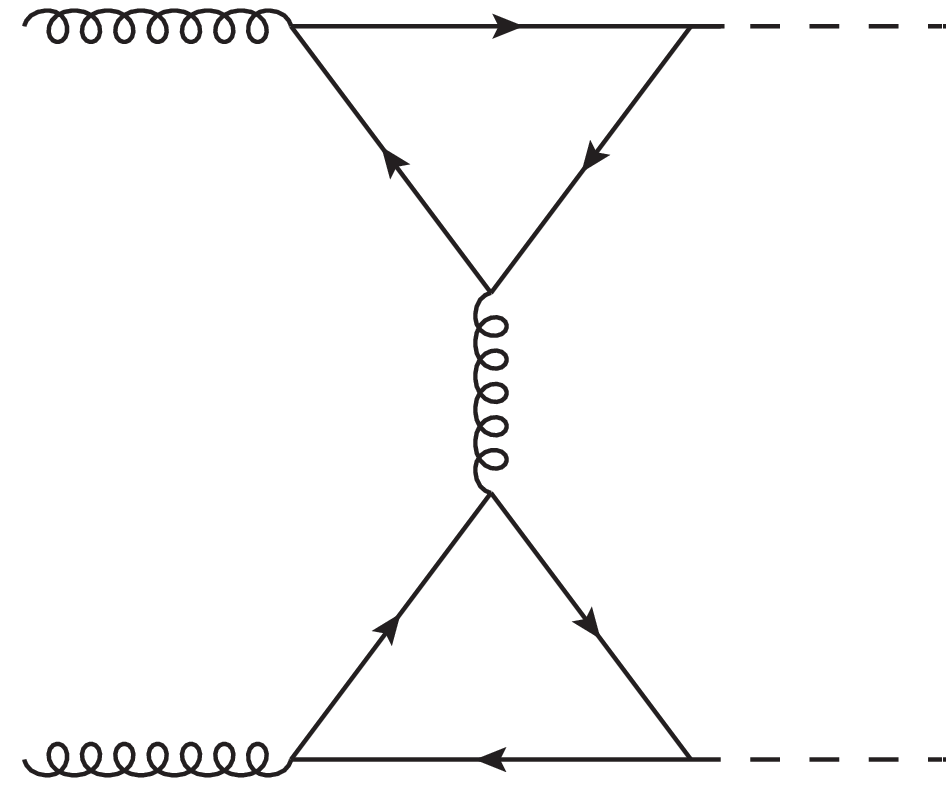}
  \includegraphics[width=.22\linewidth]{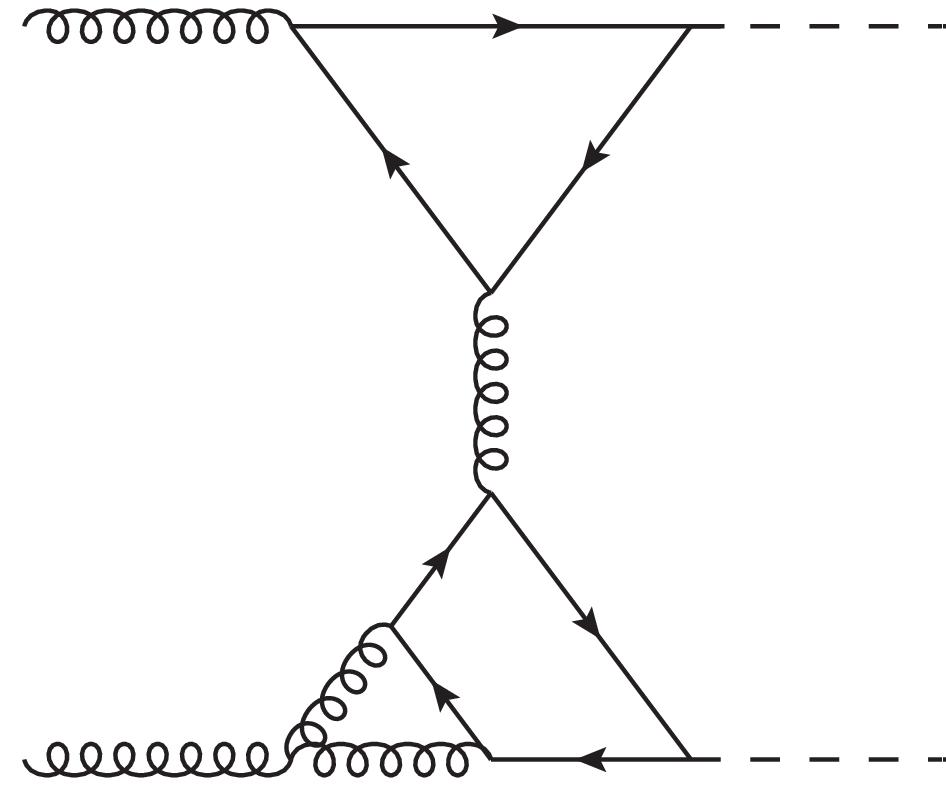}
  \includegraphics[width=.22\linewidth]{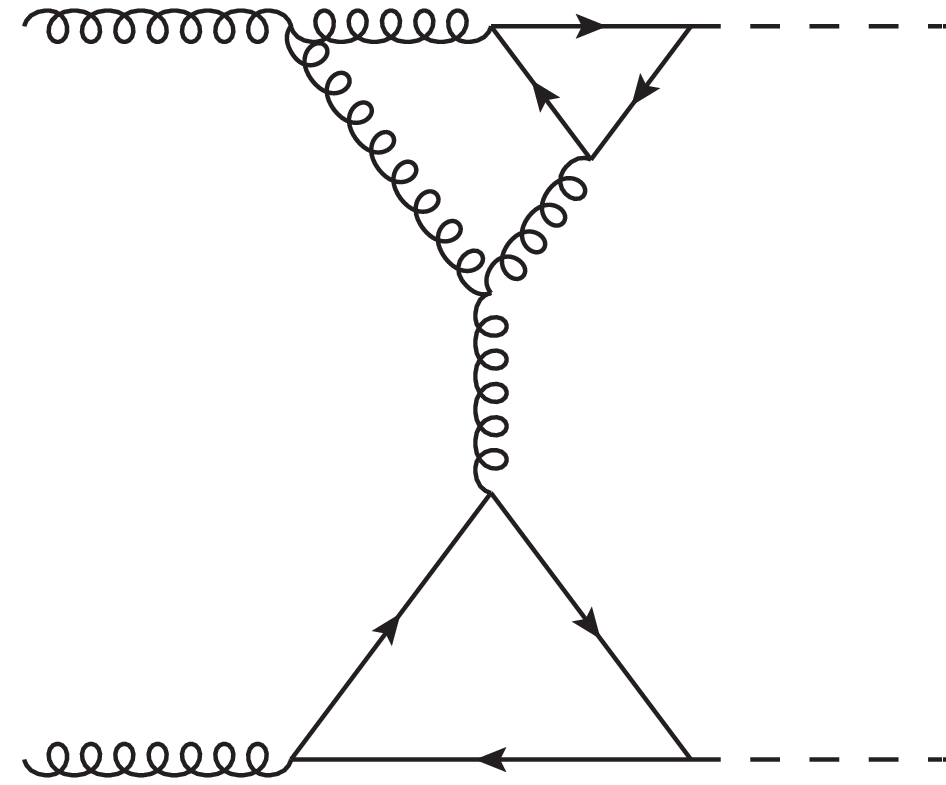}
  \includegraphics[width=.22\linewidth]{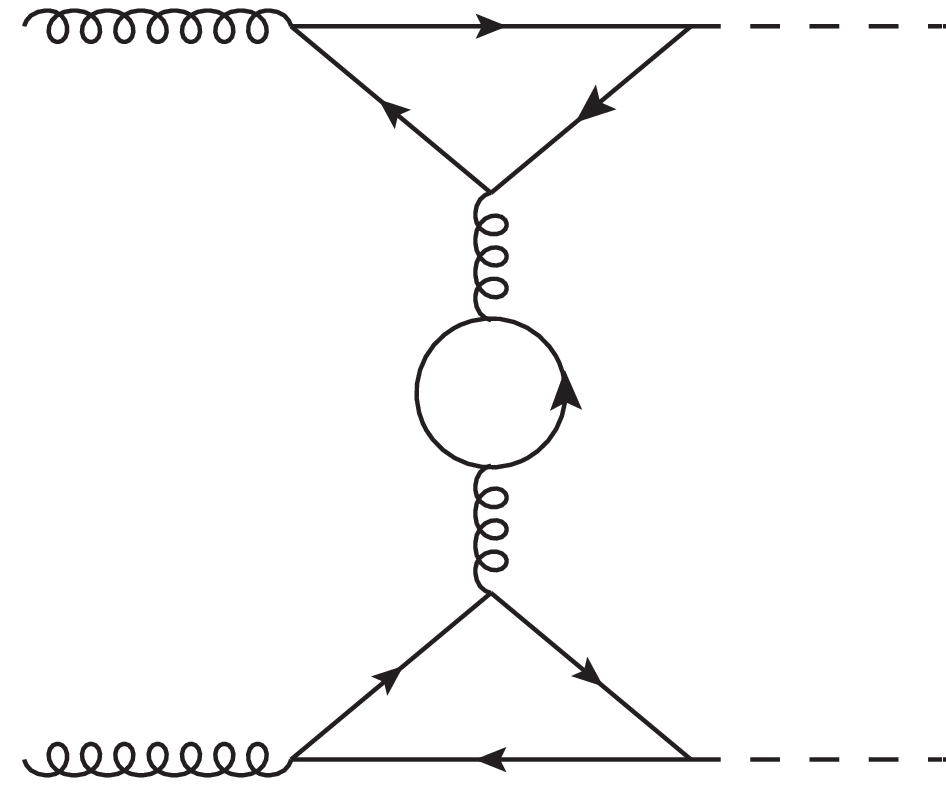}
  \caption{Sample two- and three-loop Feynman
    diagrams contributing to the one-particle reducible part of 
    $gg\to HH$. Solid, dashed and curly lines represent top quarks, Higgs bosons and gluons, respectively.}\label{fig::diags}
\end{figure}

In this paper we compute the reducible contribution to class (ii).
It can be composed from one- and two-loop corrections to the
gluon-gluon-Higgs vertex with an off-shell gluon and
one-loop corrections to the gluon propagator. Some sample
Feynman diagrams are depicted in Fig.~\ref{fig::diags}.

For completeness we briefly repeat the definition
of the form factors for $g(q_1)g(q_2)\to H(q_3)H(q_4)$, with all momenta
$q_i$ defined to be incoming (thus, $q_4=-q_1-q_2-q_3$).
The amplitude can be decomposed into two Lorentz structures
\begin{eqnarray}
  {\cal M}^{ab} &=& 
  \varepsilon_{1,\mu}\varepsilon_{2,\nu}
  {\cal M}^{\mu\nu,ab}
  \,\,=\,\,
  \varepsilon_{1,\mu}\varepsilon_{2,\nu}
  \delta^{ab} X_0 s 
  \left( F_1 A_1^{\mu\nu} + F_2 A_2^{\mu\nu} \right)
  \,,
                    \label{eq::M}
\end{eqnarray}
where $a$ and $b$ are adjoint colour indices and $s=(q_1+q_2)^2$ is the
squared partonic centre-of-mass energy.  
The two Lorentz structures are given by
\begin{eqnarray}
  A_1^{\mu\nu} &=& g^{\mu\nu} - {\frac{1}{q_{12}}q_1^\nu q_2^\mu
  }\,,\nonumber\\
  A_2^{\mu\nu} &=& g^{\mu\nu}
                   + \frac{1}{{p_T^2} q_{12}}\left(
                   q_{33}    q_1^\nu q_2^\mu
                   - 2q_{23} q_1^\nu q_3^\mu
                   - 2q_{13} q_3^\nu q_2^\mu
                   + 2q_{12} q_3^\mu q_3^\nu \right)\,,
\end{eqnarray}
with
\begin{eqnarray}
  q_{ij} &=& q_i\cdot q_j\,,\qquad
  {p_T^{\:2}} \:\:\:=\:\:\: \frac{2q_{13}q_{23}}{q_{12}}-q_{33}
  {\:\:\:=\:\:\: \frac{tu - m_H^4}{s}}
\end{eqnarray}
where $s$, ${t=(q_1+q_3)^2}$ and ${u=(q_2+q_3)^2}$ are Mandelstam variables which fulfill
$s+t+u=2m_H^2$. The quantity $X_0$ is given by
\begin{eqnarray}
  X_0 &=& \frac{G_F}{\sqrt{2}} \frac{\alpha_s(\mu)}{2\pi} T_F \,,
\end{eqnarray}
with $T_F=1/2$, $G_F$ is Fermi's constant and $\alpha_s(\mu)$ is the strong
coupling constant evaluated at the renormalization scale $\mu$.

We define the expansion in $\alpha_s$ of the form factors as
\begin{eqnarray}
  F &=& F^{(0)} + \frac{\alpha_s(\mu)}{\pi} F^{(1)} 
  + \left(\frac{\alpha_s(\mu)}{\pi}\right)^2 F^{(2)} + \cdots
  \,.
  \label{eq::F}
\end{eqnarray}
The one-particle reducible contributions which we focus on here
contribute for the first time at two loops, to $F^{(1)}$. We denote their
contribution to $F_1$ and $F_2$ by $F^{(1)}_{\rm dt1}$ and
$F^{(1)}_{\rm dt2}$, respectively.  At this order they have been
computed in Ref.~\cite{Degrassi:2016vss}, where they are given as exact
expressions in all parameters.
In this paper we reproduce
these results in an expansion for $m_H\to 0$
and add the terms of order $\epsilon$ and $\epsilon^2$ at two loops, which are necessary for
renormalization and infrared subtraction  at three loops,
and we compute the corresponding
contributions at three-loop order, $F^{(2)}_{\rm dt1}$ and
$F^{(2)}_{\rm dt2}$.  The main ingredient for this calculation are
two-loop corrections to the $gg\to H$ vertex with an off-shell gluon, as
can be seen from Fig.~\ref{fig::diags}. In the next section we
describe our approach to compute this building block in detail.
In Section \ref{sec::results} we then present our results
for the form factors $F_{\rm dt1}$ and $F_{\rm dt2}$ before we conclude in
Section \ref{sec::concl}.


\section{Calculation of the $gg\to H$ vertex at two loops}


\subsection{$gg\to H$ form factors}

In this section we describe the calculation of the 
$gg\to H$ building block, which we have to consider up to two-loop order.
We introduce the
three-point amplitude $\mathcal{V}^{\alpha \beta}(q_s,q_2)$ for the
interaction of an off-shell gluon, an on-shell gluon and a Higgs boson.
Assuming that the off-shell gluon has momentum $q_s$, the
on-shell gluon $q_2$ and that all momenta are incoming we have
$(q_s+q_2)^2 = m_H^2$ and the most general
Lorentz decomposition of the amplitude is
\begin{equation}
  \mathcal{V}^{\alpha \beta}(q_s,q_2) = 
  V_0\left(
  F_a ~ g^{\alpha \beta} (q_s \cdot q_2) 
  + F_b ~ q_s^\alpha q_2^\beta 
  + F_c ~ q_2^\alpha q_s^\beta 
  + F_d ~ q_s^\alpha q_s^\beta 
  + F_e ~ q_2^\alpha q_2^\beta
  \right)\,,
\label{eq:gghlorentz}
\end{equation}
where the form factors $F_x$ are dimensionless. For the expansion in $\alpha_s$ we follow Eq.~(\ref{eq::F}). The normalization factor $V_0$ is chosen such that 
\begin{eqnarray}
   F_a^{(0)}(s=0,m_H=0) &=& \frac{4}{3}\,. 
\end{eqnarray}
Furthermore, we decompose $F_x^{(1)}$ according to the SU($N_c$) colour factors $C_A=N_c$ and $C_F=(N_c^2-1)/(2N_c)$
\begin{eqnarray}
  F_x^{(1)} &=& C_A F_x^{(1),C_A} + C_F F_x^{(1),C_F}\,.
\end{eqnarray}
Note that $F_a,\ldots,F_e$ are
not all independent. Typically, imposing the Ward
identities allows one to derive relations among them. However, since
one gluon is off-shell, more structures  give a
non-zero contribution compared to the case in which both gluons are
on-shell. The Ward identities in our case are obtained by
the simultaneous contraction of both gluon momenta with the
amplitude,
\begin{equation}
q_{s,\alpha} q_{2,\beta} \mathcal{V}^{\alpha \beta}(q_s,q_2) = 0,
\end{equation}
which leads to the following relation between the form factors 
\begin{equation}
F_d = -\frac{(F_a + F_c) (m_H^2-q_s^2)}{2 q_s^2}  \,.
\label{eq:facdwi}
\end{equation}
For the calculation of the form factors for Higgs pair production $F_b$ and $F_e$ are not needed.
We note that at one-loop order we find $F_d^{(0)}=0$, so that the above relation
implies that $F_c^{(0)} = - F_a^{(0)}$,
as in the case of two on-shell gluons.  At two-loop
order, however, $F_d^{(1)}$ is not zero; we have verified that our results satisfy Eq.~\eqref{eq:facdwi}. At
two-loop order $F_d^{(1)}$ only has a contribution proportional to the colour
factor $C_A$.

The one-loop form factor $F_a^{(0)}$ 
is finite and gauge-parameter
independent. At two-loop order the form factors depend on the QCD
gauge parameter $\xi$ and furthermore develop $1/\epsilon$ poles,
which have both ultraviolet and infrared origin. The $\xi$ dependence
and $1/\epsilon$ poles are also present in the three-loop one-particle
reducible parts of the Higgs boson pair production form factors $F_{\rm dt1}$
and $F_{\rm dt2}$. Here the $\xi$ dependence cancels after the combination
with the contribution (b) from Fig.~\ref{fig::diag_gghh}, as
we have shown previously in Ref.~\cite{Davies:2019djw} in the context of the large top quark mass expansion.
The cancellation of the ultraviolet $1/\epsilon$ poles requires additionally the renormalization of the top quark mass, the strong coupling constant and the gluon wave function. The infrared poles must be subtracted using an
appropriate prescription (see, e.g., Ref.~\cite{Catani:1998bh}), or the
real-radiation contributions must be added.
In this paper we  present bare results, both for the
$gg\to H$ vertex and the $gg \to HH$ form factors $F_{\rm dt1}$ and $F_{\rm dt2}$, with
explicit dependence on $\xi$ and $\epsilon$.


\subsection{Workflow}

In the following we describe the workflow for the computation
of the $gg\to H$ vertex. As described above, we give the off-shell gluon
the momentum $q_s$. When we use the resulting building block to construct
$F_{\rm dt1}$ and $F_{\rm dt2}$ for $gg\to HH$, we will have either $q_s^2=t$ or $q_s^2=u$.
Thus, the relevant kinematic region for the $gg\to H$ vertex has $q_s^2<0$.

We first generate the amplitude in terms of Lorentz-scalar functions
using our usual chain of programs ({\tt qgraf}~\cite{Nogueira:1991ex}, {\tt
  tapir}~\cite{Gerlach:2022qnc}, {\tt
  exp}~\cite{Harlander:1998cmq,Seidensticker:1999bb} and the in-house {\tt
  FORM}~\cite{Ruijl:2017dtg} code ``{\tt calc}'').  We then  perform the
integration-by-parts reduction to master integrals using
{\tt Kira}~\cite{Klappert:2020nbg}, arriving at 4 and 46 master integrals
at one and two loops, respectively. Up to this point, we retain exact
dependence on the kinematic parameters $q_s^2$, $m_H$ and $m_t$.

The next step is to compute the master integrals. Here we distinguish two
kinematic regions, which allows us to obtain compact (semi-)analytic
results for the form factors. In the first region, we expand the master
integrals around $m_H\to 0$, which leads to a good approximation for larger
values of $|q_s^2|$. For smaller values we instead expand
for $|q_s^2|\to 0$. In the following two sub-sections, we describe our
approach for each region.


\subsection{Small $m_H$ expansion}\label{sec:smallmhexp}

\begin{figure}[t]
  \centering
    \includegraphics[width=0.36\linewidth]{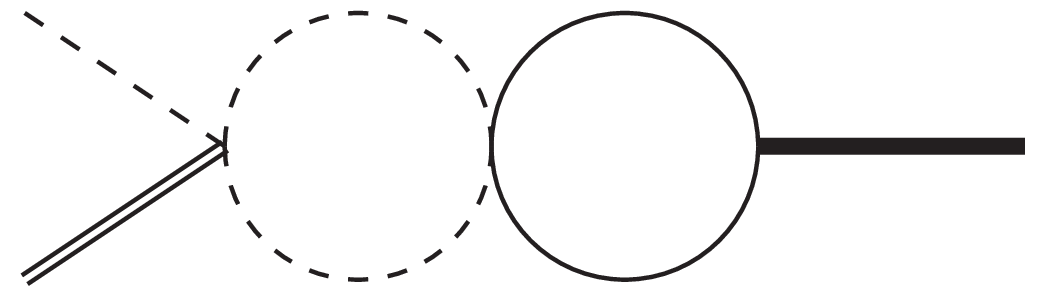}
    \includegraphics[width=0.26\linewidth]{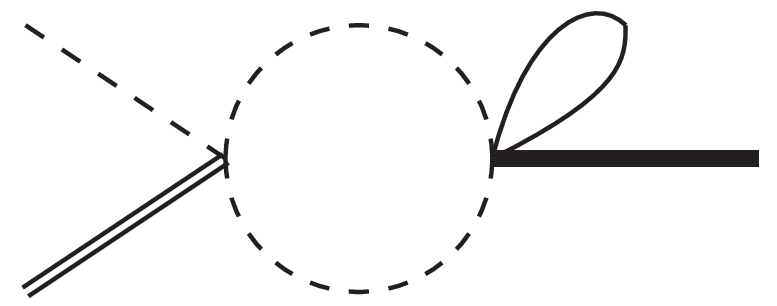}
    \includegraphics[width=0.36\linewidth]{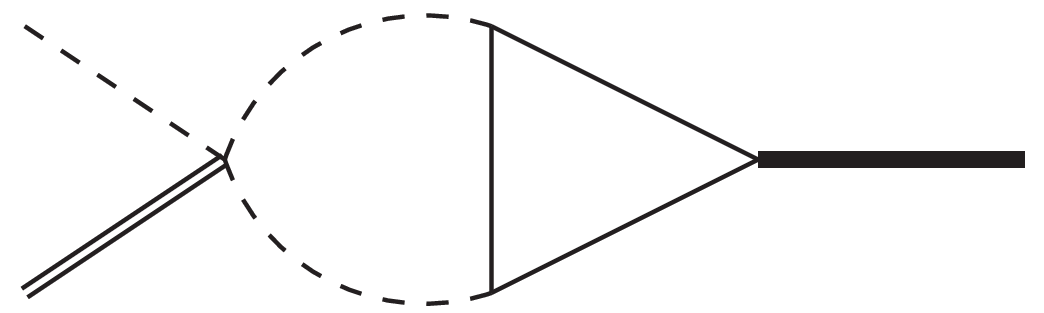}
  \caption{The three out of the 46 two-loop  master integrals which have a soft and hard contribution. Dashed lines denote massless internal and external legs, while solid thin lines are used for massive internal legs with mass $m_t$. Double lines and solid thick lines denote massive external legs with virtuality $q_s^2$ and mass $m_H$, respectively.}\label{fig::softMI}
\end{figure}

At one-loop order the expansion of the master integrals for small $m_H$ is a
Taylor expansion, which is conveniently realized with the help of {\tt
  LiteRed}~\cite{Lee:2013mka}.  At two-loop order one must consider two
regions; in the hard region we can again use {\tt LiteRed} for a Taylor
expansion in $m_H$.  The second region arises from diagrams which have
an external momentum $(q_s+q_2)^2=m_H^2$ and at the same time a cut
through gluon lines. Among the 46 two-loop master integrals there are three
such integrals, which are shown in Fig.~\ref{fig::softMI}. None of them depend on $q_s^2$, thus
the expansion for $m_H\to0$ is equivalent to the large-$m_t$
expansion which is straightforward to perform using ${\tt exp}$.

After the expansion in $m_H$, in the hard region we obtain new integral
families. For the integration-by-parts reduction we use again {\tt
  Kira} and obtain 25 master integrals. Their internal lines are
either massless or have mass $m_t$. Two of the external lines are
massless and one has the off-shell squared momentum $q_s^2$;
the master integrals depend only on the ratio $q_s^2/m_t^2$.

The one-loop master integrals can be computed analytically in terms of
harmonic polylogarithms~\cite{Remiddi:1999ew}.  At two-loop order we use the ``expand and match''
approach~\cite{Fael:2021kyg,Fael:2022rgm,Fael:2022miw,Fael:2023zqr}
which uses the differential equations for the master integrals to
construct deep generalized expansions around properly chosen values of
$q_s^2/m_t^2$, with high-precision numerical coefficients. The boundary
values at $q_s^2/m_t^2 \to 0$ are obtained analytically using the
large-$m_t$ expansion, and then transported numerically to the other
expansion points with high precision.

Using this approach we obtain expansions
for each master integral up to order $m_H^8$,
where the coefficients are ``semi-analytic'' piecewise functions
of $q_s^2/m_t^2$. Thus a flexible implementation in a computer
code is possible, which allows for a straightforward modification of all
input parameters.


\subsection{\label{sub::smalls}Low energy expansion}

In order to cover the small $|q_s^2|$ region, we instead expand the master
integrals in the limit $q_s^2 \to 0$.
Note that this expansion is only required at the two-loop level;
at one loop, the small $m_H$ expansion covers the full kinematic
region because the external gluons couple only to massive top quarks.

The expansion for $q_s^2 \to 0$ is also an asymptotic expansion with two
regions which we denote as ``hard'' and ``soft'' in the following.
We construct the expansion from the differential equation by 
separating the two branches of the asymptotic expansion and 
deriving power series solutions, as is usually done in the 
``expand and match'' approach.
In the low energy expansion it is again possible to
fix the boundary values analytically.

The expansion in the hard region can be realized by a Taylor series in the 
off-shell momentum $q_s$. 
The resulting integrals are the same as for on-shell 
Higgs production at two-loop order~\cite{Spira:1995rr,Harlander:2005rq,Anastasiou:2006hc,Aglietti:2006tp,Harlander:2009bw}.
We have re-derived the solutions by solving the system of differential 
equations utilizing the algorithm of Ref.~\cite{Ablinger:2018zwz} implemented 
with the help of the packages \texttt{HarmonicSums} \cite{Blumlein:1998if,Vermaseren:1998uu,Blumlein:2009ta,Ablinger:2009ovq,Ablinger:2011te,Ablinger:2012ufz,Ablinger:2013eba,Ablinger:2013cf,Ablinger:2014bra,Ablinger:2014rba,Ablinger:2015gdg,Ablinger:2018cja} 
and \texttt{Sigma} \cite{sigmaI,sigmaII}.
The only boundary condition necessary is the two-loop tadpole;
the results can be expressed in terms of harmonic polylogarithms 
\cite{Remiddi:1999ew} of argument $x_{m_H}$ defined by
\begin{align}
    \frac{m_H^2}{m_t^2} &= -\frac{(1-x_{m_H})^2}{x_{m_H}} ~.
    \label{eq::trafo}
\end{align}

The boundary conditions in the soft region are obtained by direct 
integration and summation techniques.
We first reveal the scaling of the $\alpha$-parameters 
of the Schwinger parameterization of the Feynman integrals with the 
help of \texttt{asy} \cite{Jantzen:2012mw} and integrate them in terms of a one-dimensional 
Mellin-Barnes representation.
The remaining integral is solved by summing residues symbolically with 
\texttt{EvaluateMultiSums} \cite{RISC4711,RISC4826}.
The final step is to convert the infinite sums into a representation 
in terms of harmonic polylogarithms, using \texttt{HarmonicSums}.

Let us demonstrate the steps explicitly for a simple example.
One of the master integrals which we have to calculate in the limit $q_s^2 \to 0$ is given by
\begin{align}
    I_{21} &= 
    \int \frac{{\rm d}^d k_1}{(2\pi)^d}
    \int \frac{{\rm d}^d k_2}{(2\pi)^d}
    \frac{1}{[m_t^2-k_1^2][m_t^2-(k_2-q_s-q_2)^2]
    [-(k_1-k_2+q_s)^2][-(k_1-k_2)^2]}~.
\end{align}
\texttt{asy} finds the soft region 
$I=\{0,0,-1,-1\}$ and the hard region 
$I_h=\{0,0,0,0\}$. 
In the soft region we obtain the one-dimensional 
Mellin-Barnes representation 
\begin{align}
    I_{21}^{(I)} &= 
    \frac{1}{2\pi i} \, \mathcal{N} \, \int\limits_{-i\infty}^{+i\infty} 
    {\rm d}\sigma (-\rho)^{\sigma}
    \frac{\Gamma(1-\epsilon)\Gamma(\epsilon)
    \Gamma(-\sigma)\Gamma^2(1+\sigma)
    \Gamma(1-\epsilon+\sigma)\Gamma(\epsilon+\sigma)}{
    \Gamma(2-2\epsilon+\sigma)\Gamma(2+2\sigma)
    } ~,
\end{align}
with $\rho=m_H^2/m_t^2$ and 
$\mathcal{N}=(-s/m_t^2)^{-\epsilon}
\exp(-2\epsilon \gamma_E)$.
By closing the integration contour to the right and using Cauchy's residue theorem we can turn the 
integration into an infinite sum
\begin{align} 
    I_{21}^{(I)} &=
    \mathcal{N} \,
    \sum\limits_{k=0}^{\infty} \rho^k
    \Gamma(1-\epsilon)\Gamma(\epsilon) 
    \frac{\Gamma(1+k)\Gamma(1-\epsilon+k)\Gamma(\epsilon+k)}{\Gamma(2-2\epsilon+k)\Gamma(2+2k)}
    \\ 
    &= \left(\frac{-s}{m_t^2}\right)^{-\epsilon}
    \!\!\biggl\{ 
    \frac{1}{\epsilon^2}
    + \frac{1}{\epsilon}
    \biggl[ 
    \frac{1}{16} \big(32\!+\!44 \rho \!-\!12 \rho^2\!+\!\rho^3\big)
    -\frac{\left(\rho(4\!-\!\rho)\right)^{3/2}}{4\rho^2} G_1
    -\frac{1}{4 \rho} G_1^2
    \biggr]
    \!+\! \mathcal{O}(\epsilon^0)
    \biggr\}\nonumber
    ,
\end{align}
with 
\begin{align} 
    G_1 &= \int\limits_{0}^{\rho} 
    {\rm d}t \sqrt{t}\sqrt{4-t}\,.
\end{align}
The symbolic summation has been 
performed with \texttt{EvaluateMultiSums} and 
\texttt{HarmonicSums} and for brevity we do not show 
higher terms in the $\epsilon$-expansion here,
though they are needed for the full calculation. 
Finally we can transform to the variable 
defined in Eq.~\eqref{eq::trafo} to obtain
\begin{align} 
    I_{21}^{(I)} &= \mathcal{N} \,
    \biggl\{
    \frac{1}{\epsilon^2}
    + \frac{1}{\epsilon}
    \biggl[
        5 
        + \frac{1+x_{m_H}}{1-x_{m_H}}H_0(x_{m_H}) 
        - \frac{x_{m_H}}{(1-x_{m_H})^2} H_0^2(x_{m_H})
    \biggr]
    + \mathcal{O}(\epsilon^0)
    \biggr\}~.
\end{align}

The final results for the master integrals are obtained by summing the
contributions of both asymptotic regions.  Finally, after inserting the master
integrals into the expanded amplitude, we obtain a consistent power-log expansion (we compute 19 terms) in
the limit $q_s^2 \to 0$ of the form factors where the coefficients are given
by analytic expressions depending on the variable $x_{m_H}$ and only harmonic
polylogarithms are necessary.  Therefore, evaluating the amplitude for small
values of $q_s^2$ is fast.

We stress here that although we have fully analytic results for the
low energy expansion, the small $m_H$ expansion of Section \ref{sec:smallmhexp}
is ``semi-analytic''.


\subsection{One-loop results for the $gg\to H$ vertex}

Let us start with the discussion of the one-loop result for the $gg\to H$ vertex
with an off-shell gluon, which we need up to order $\epsilon$. Here the
expansion in $m_H$ works very well, even for small values of $|q_s^2|$. For
convenience we present explicit results for\footnote{Note that the other form
factors are either zero, or are not needed for $F_{\rm dt1}$ and $F_{\rm dt2}$.}
$F_a^{(0)}$ including terms up to
${\cal O}(m_H^2)$ and ${\cal O}(\epsilon)$.
We set $\mu=m_t$ for brevity.
Expansions
up to order $m_H^4$ and $\epsilon^2$ and with full 
dependence on the renormalization scale
can be found in the ancillary files of this
paper\footnote{Terms beyond $m_H^4$ are very small for all $q_s^2$
values.}. Our result reads 
\begin{eqnarray}
  F_a^{(0)} &=& 
    \frac{24 x}{(1-x)^2}
    +\frac{8 x (1+x) H_0}{(1-x)^3}
    +\frac{4 x \big(1-6 x+x^2\big) H_{0,0}}{(1-x)^4}
    + \epsilon \biggl[ 
        \frac{56 x}{(1-x)^2}
        +\frac{24 x (1+x) H_0}{(1-x)^3}
        \nonumber \\ &&
        +\frac{8 x \big(1-2 x-x^2\big) H_{0,0}}{(1-x)^4}
        -\frac{16 x (1+x) H_{-1,0}}{(1-x)^3}
        +\frac{4 x \big(1-6 x+x^2\big) H_{0,0,0}}{(1-x)^4}
        \nonumber \\ &&
        -\frac{8 x \big(1-6 x+x^2\big) H_{0,-1,0}}{(1-x)^4}
        -\frac{12 x \big(1-6 x+x^2\big) \zeta (3)}{(1-x)^4}
        -\pi ^2 
        \biggl(
             \frac{4 x (1+x)}{3 (1-x)^3}
             \nonumber \\ &&
            +\frac{2 x \big(1-6 x+x^2\big) H_0}{3 (1-x)^4}
        \biggr)
    \biggr]
    + \frac{m_H^2}{m_t^2}
    \Biggl\{
    \frac{2 x \big(1-74 x+x^2\big)}{3 (1-x)^4}
    -\frac{16 x^2 (1+x) H_0}{(1-x)^5}
    \nonumber \\ &&
    -\frac{4 x^2 \big(1-10 x+x^2\big) H_{0,0}}{(1-x)^6}
    + \epsilon \biggl[ 
        -\frac{112 x^2}{(1-x)^4}
        -\frac{48 x^2 (1+x) H_0}{(1-x)^5}
        +\frac{32 x^2 (1+x) H_{-1,0}}{(1-x)^5}
        \nonumber \\ &&
        -\frac{16 x^2 \big(1-2 x-x^2\big) H_{0,0}}{(1-x)^6}
        -\frac{4 x^2 \big(1-10 x+x^2\big) H_{0,0,0}}{(1-x)^6}
        +\frac{8 x^2 \big(1-10 x+x^2\big) H_{0,-1,0}}{(1-x)^6}
        \nonumber \\ &&
        +\pi ^2 
        \biggl(
             \frac{8 x^2 (1+x)}{3 (1-x)^5}
            +\frac{2 x^2 \big(1-10 x+x^2\big) H_0}{3 (1-x)^6}
        \biggr)
        +\frac{12 x^2 \big(1-10 x+x^2\big) \zeta_3}{(1-x)^6}
    \biggr]
    \Biggr\}
    \nonumber \\ &&
    +{\cal O}(m_H^4) +{\cal O}(\epsilon^2)\,,
\end{eqnarray}
with 
$q_s^2/m_t^2 = -(1-x)^2/x$
and $H_{\vec{w}}\equiv H_{\vec{w}}(x)$.

\begin{figure}[tbh]
  \centering
  \begin{tabular}{ccc}
    \includegraphics[width=.5\linewidth]{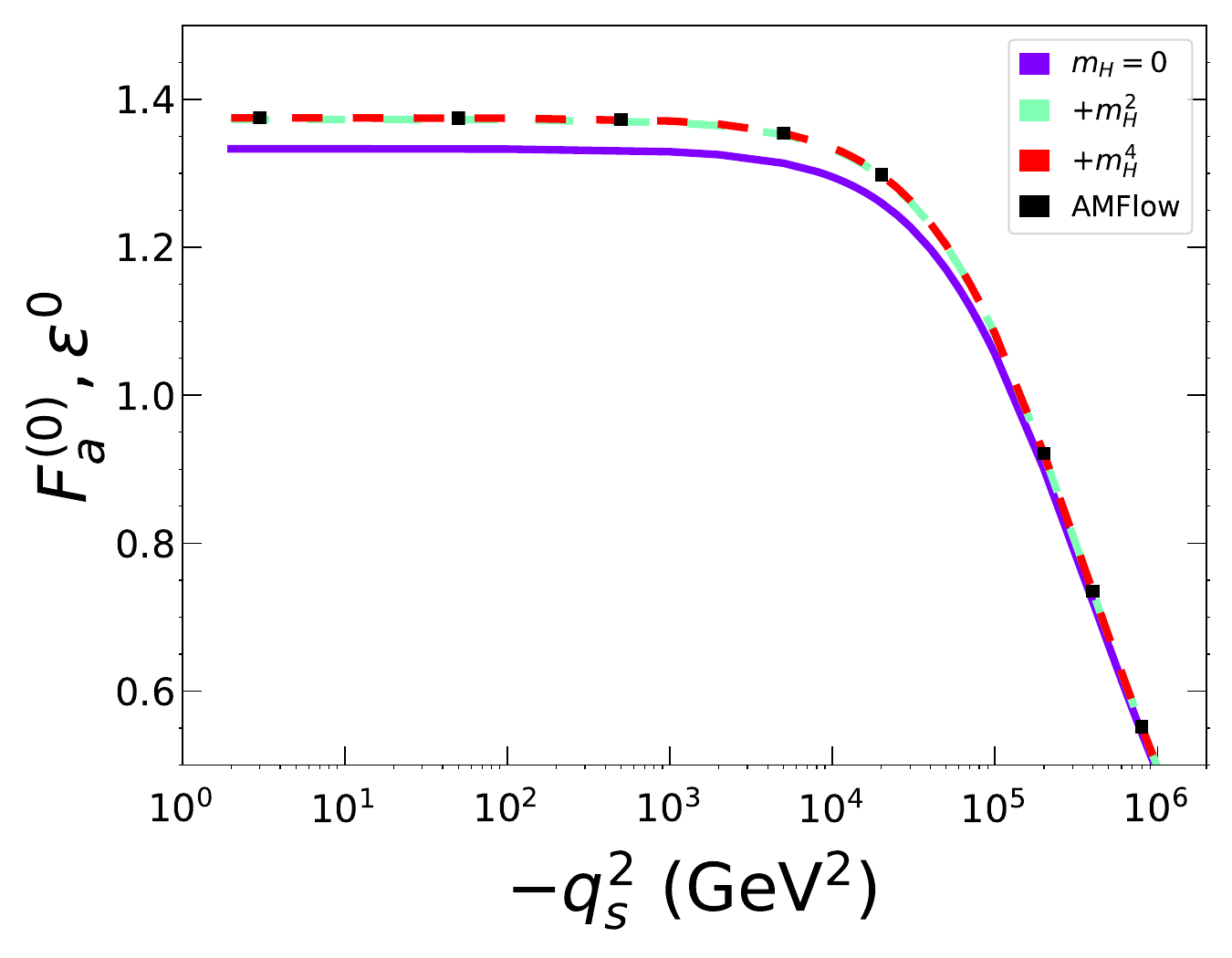} & \includegraphics[width=.5\linewidth]{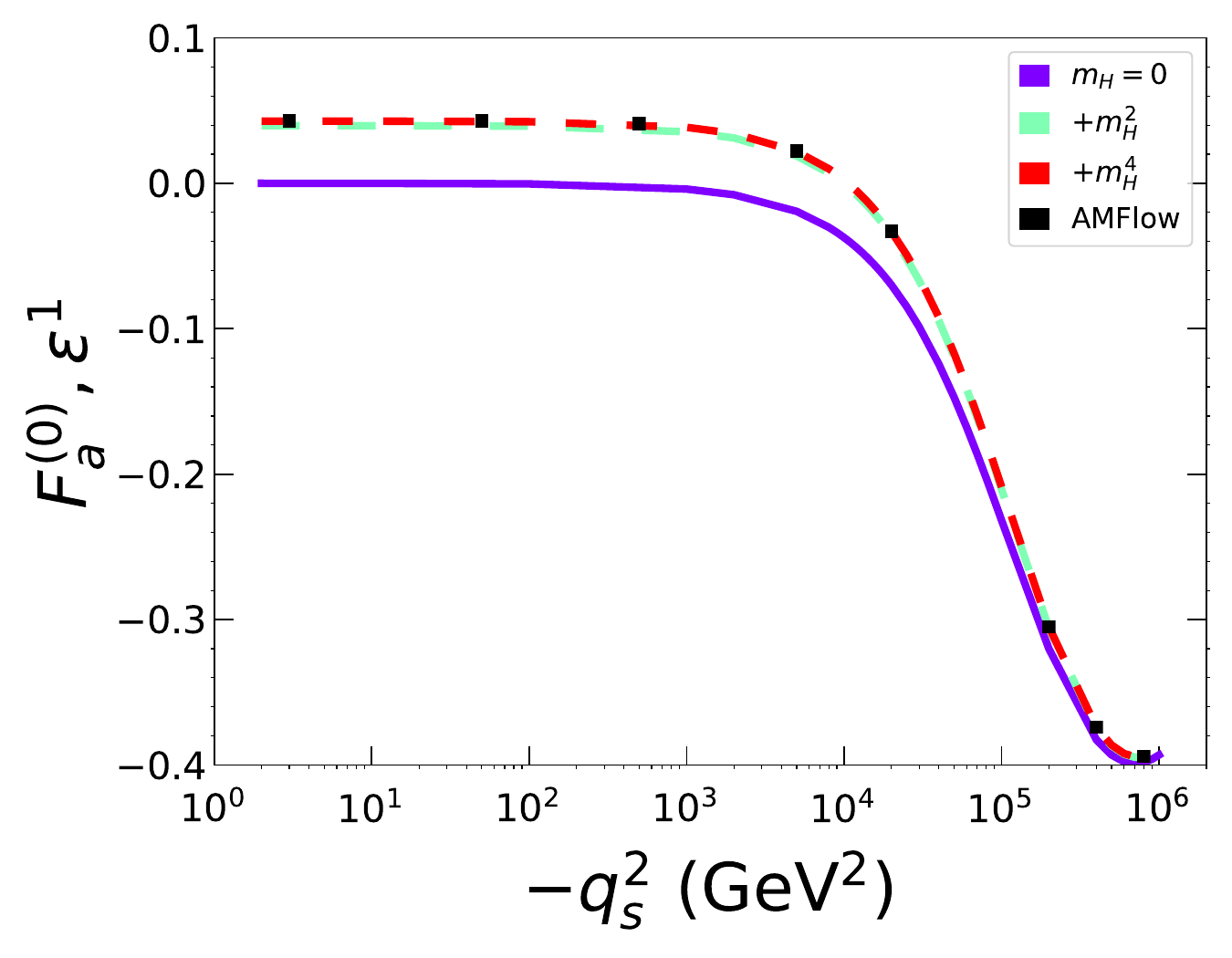}
  \end{tabular}
  \caption{$F_a^{(0)}$ as a function of $q_s^2$. Both the
  $\epsilon^0$ (left) and $\epsilon^1$ terms (right) are shown.
  }\label{fig::Fa0}
\end{figure}

In Figure~\ref{fig::Fa0} we show $F_a^{(0)}$ as a function of $q_s^2$, at
several expansion depths in $m_H$.
Furthermore we show results obtained
from the un-expanded amplitude, with the four master integrals evaluated numerically using
{\tt AMFlow}~\cite{Liu:2022chg}.  For the values for the top
quark and Higgs boson masses we use
\begin{eqnarray}
  m_t = 175~\mbox{GeV} &\mbox{and}& m_H = 125~\mbox{GeV}\,.
\end{eqnarray}
We observe a rapid convergence of the $m_H$ expansion, and very good
agreement with the numerical results. 
For example, for $q_s^2=-3$~GeV$^2$ the deviation of the $m_H^4$ expansion
from the numerical results is about 0.01\%, 0.7\% and 0.02\% for the $\epsilon^0$, $\epsilon^1$ and
$\epsilon^2$ terms, respectively.
For all practical purposes it is
sufficient to work with the small $m_H$ approximation, including
terms up to order $m_H^4$; these expressions are much more convenient to
work with, since only simple harmonic polylogarithms are present.

\subsection{Two-loop results for the $gg\to H$ vertex}

We first want to discuss the quality of our approximations.
In the case of the low energy expansion we use the master integrals
from Section \ref{sub::smalls}, insert them into the amplitude and
expand the whole amplitude consistently in $q_s^2$. This leads to compact
expressions.

If $|q_s^2|$ is sufficiently large we can construct a consistent
expansion in $m_H$, i.e., after inserting the $m_H$-expanded master integrals
into the amplitude we expand the whole expression in $m_H$, up to order
$m_H^8$. In the plots the corresponding curves have the label ``$m_H^{\rm
  EXP}\to0$''. 
However, as we will see below, in the intermediate $q_s^2$ region,
$100~\mbox{GeV} \lesssim \sqrt{|q_s^2|} \lesssim 170$~GeV,
it is crucial to \emph{not} expand the coefficients of the master integrals in
$m_H$, in order to obtain stable numerical results for $F_a^{(1),C_A}$.
In the plots
the corresponding curves are denoted by ``$m_H\to0$''.
The form factor $F_a^{(1),C_F}$ does not have this problem, since there are no
contributing Feynman diagrams with a massless cut. As a consequence,
$F_a^{(1),C_F}$ can be treated in the same way as the one-loop form factor,
i.e., the consistent
$m_H$ expansion covers the whole phase space.
This we also observe for
$F_d^{(1)}$ although it 
obtains contributions from diagrams with massless cuts; see discussion below.

Note that in  the latter case where we use $m_H$-expanded master integrals
but do not expand their coefficients,
there is no analytic cancellation of the spurious higher-order $\epsilon$ poles,
which
in this case are $1/\epsilon^2$ and $1/\epsilon^3$.  However, their coefficients
are small, typically four to five orders of magnitude smaller than the
coefficient at order $\epsilon^0$; we take this to be a measure of our
numerical accuracy.

\begin{figure}[tb]
  \centering
  \begin{tabular}{cc}
    \includegraphics[width=.45\linewidth]{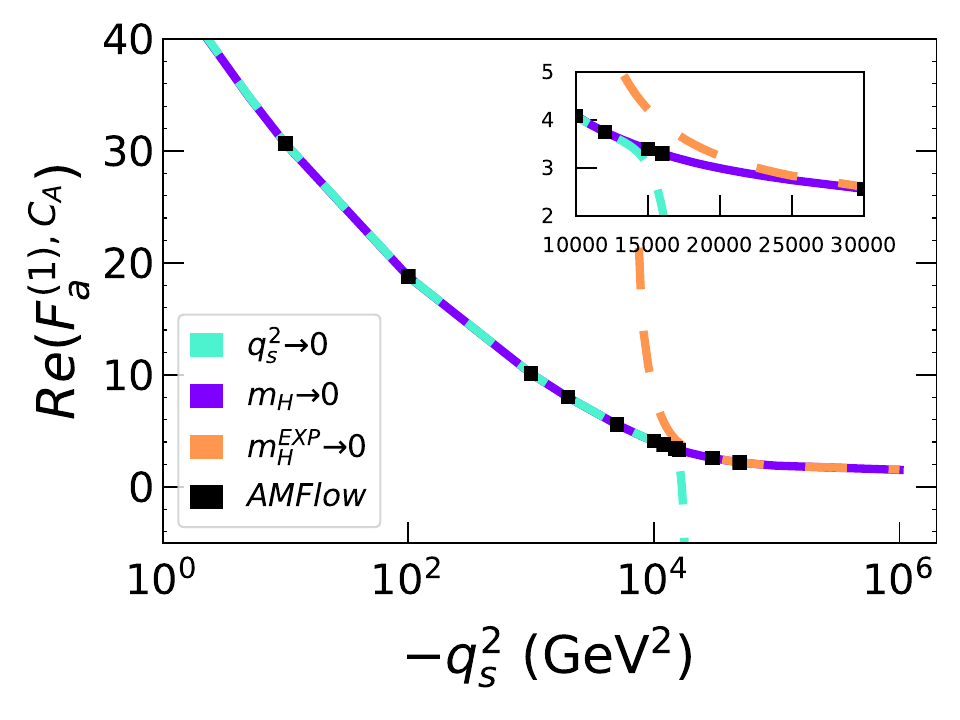} & 
    \includegraphics[width=.45\linewidth]{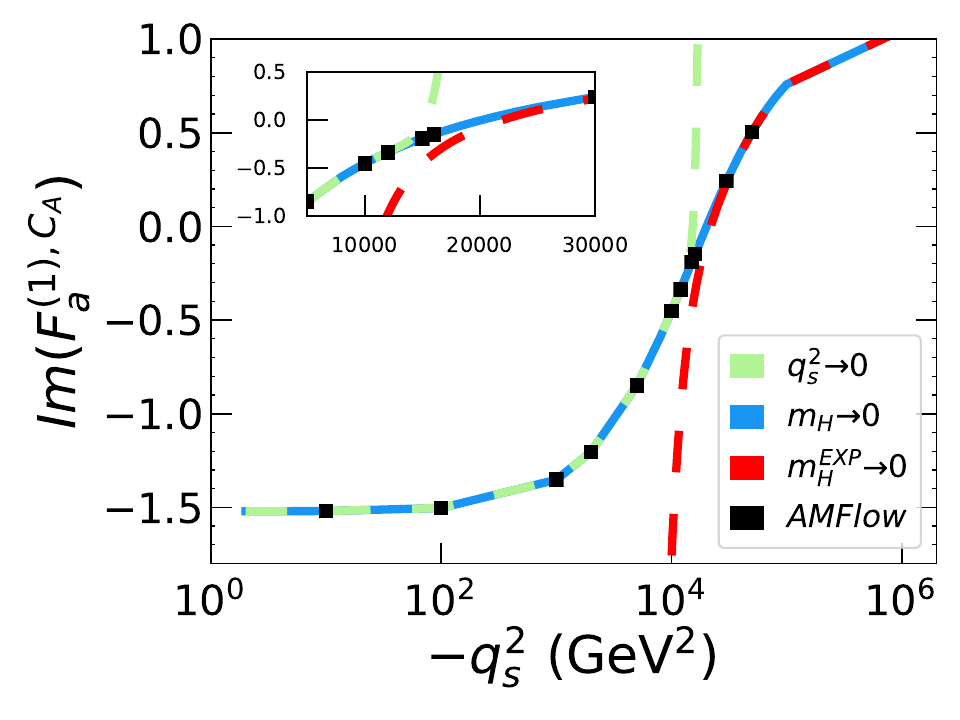}
    \\
    \includegraphics[width=.45\linewidth]{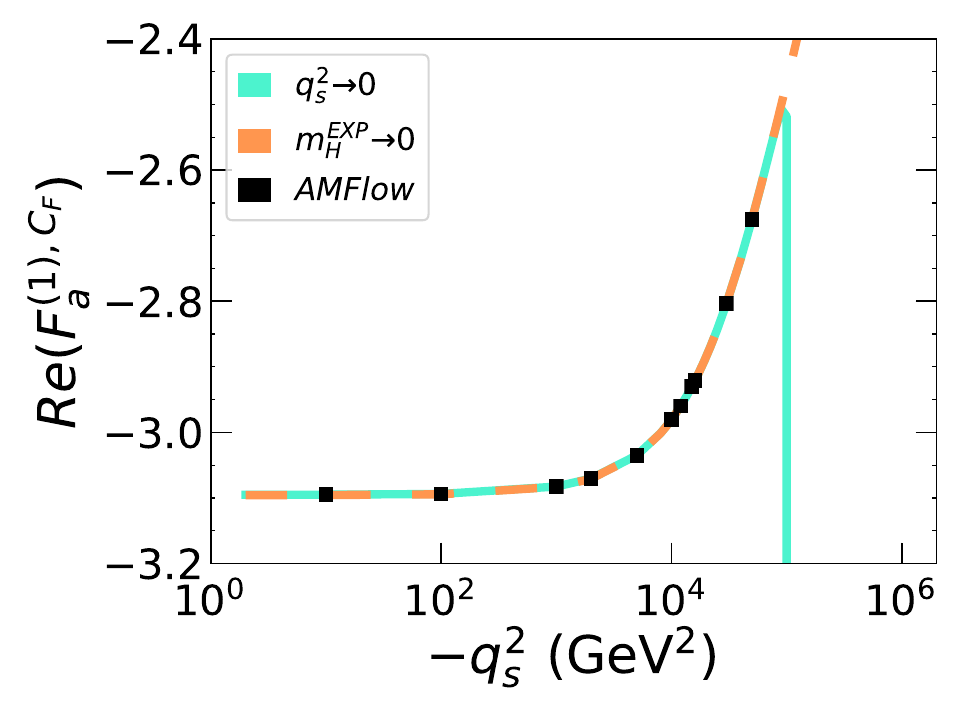} &
    \includegraphics[width=.45\linewidth]{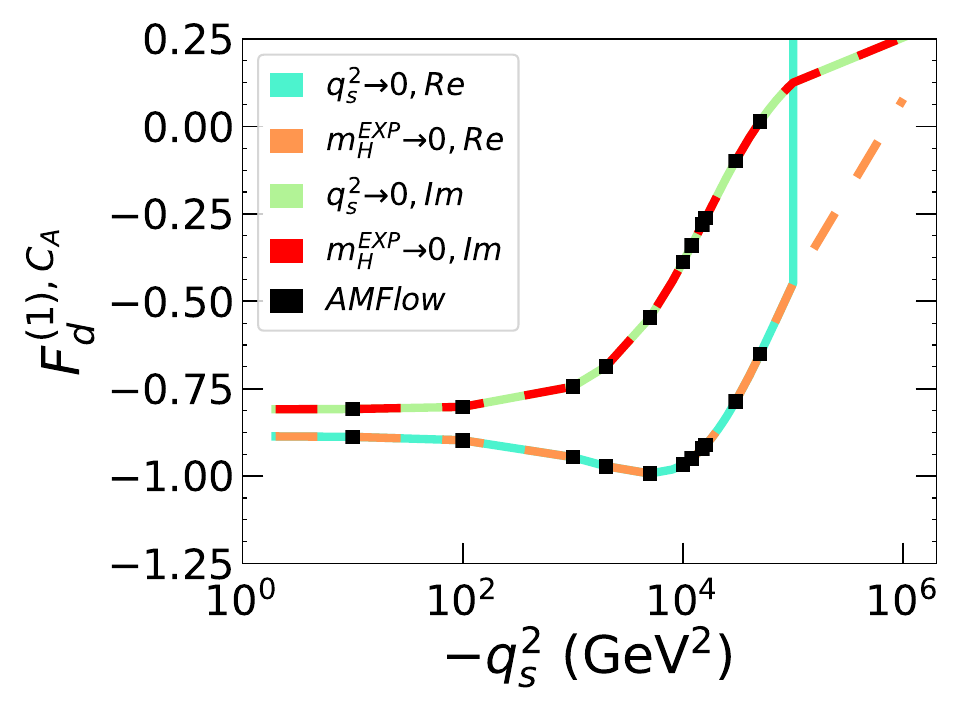}
  \end{tabular}
  \caption{$F_a^{(1)}$ and $F_d^{(1)}$ as a function of
    $q_s^2$. }
  \label{fig::Facd1}
\end{figure}

In Fig.~\ref{fig::Facd1} we show the finite parts of the two-loop
form factors $F_a^{(1)}$ and $F_d^{(1)}$ as a function of
$q_s^2$. 
For the renormalization scale we have chosen $\mu^2=m_t^2$.
For convenience we choose a logarithmic scale for the $q_s^2$ axis which
enhances the region for small values of $|q_s^2|$. 
The squares in Fig.~\ref{fig::Facd1} correspond to exact results
where numerical values for the 46 master integrals have been obtained with
{\tt AMFlow}~\cite{Liu:2022chg}. The excellent agreement with
our results justifies our approximations in the various $q_s^2$ regions.

For $F_a^{(1),C_A}$ we show the
approximation for small $|q_s^2|$ and the two versions of the 
small-$m_H$ expansion (as described above). We observe
that there is a small gap, close to $-q_s^2\approx 10^4$~GeV$^2$,
between the ``$q_s^2\to0$'' and ``$m_H^{\rm EXP}\to 0$'' expansions (see inset plot). It is
here that the ``$m_H\to0$'' expansion must be used.
For $F_a^{(1),C_F}$ and $F_d^{(1),C_A}$ we only show the
``$q_s^2\to0$'' and ``$m_H\to 0$'' curves and observe that the
$m_H$-expanded approximation covers the whole $q_s^2$ range.
For $F_a^{(1),C_F}$ the two approximations agree far below the per cent level for $|q_s^2|\lesssim 90\,000$~GeV$^2$. 
Also for $F_d^{(1),C_A}$ 
one observes very good
agreement of the consistent
$m_H$ expansion with the results from the $q_s$-expansion and the numerical results (black squares).
It is interesting to note that the $q_s$-expansion also works well up to quite high values of $|q_s^2|$. In the case of the imaginary part it shows good agreement with the $m_H$ expansion even for $|q_s^2|\approx 10^6$~GeV$^2$.
The agreement is below 0.01\% for $|q_s^2|\lesssim 100\,000$~GeV$^2$ both for the real and imaginary part.

Thus, our scheme for the numerical evaluation of $F_a^{(1),C_A}$ is
as follows: For $|q_s^2|\,\lesssim\,5\,000$~GeV$^2$ we use the $q_s^2$-expansion,
for $5\,000$~GeV$^2 \,\lesssim\,|q_s^2|\,\lesssim\,80\,000$~GeV$^2$ we use the 
``$m_H\to0$'' expansion, and for
$|q_s^2|\,\gtrsim\,80\,000$~GeV$^2$ we use the ``$m_H^{\rm EXP}\to0$'' expansion.

At this point a comment is in order on the relevant values for $|q_s^2|$, once
the $gg\to H$ vertex is used to compute the $gg\to HH$ form factors $F_{\rm dt1}$ and $F_{\rm dt2}$.
In the case of double Higgs production,
if we consider e.g.~$q_s^2=t$, the ``$m_H^{\rm EXP}\to0$'' expansion is sufficient in order to cover the phase-space region where $p_T \gtrsim 280$~GeV, for every value of $\sqrt{s}$. Instead, the ``$m_H\to0$'' and the ``$m_H^{\rm EXP}\to0$'' expansions are required for $p_T \gtrsim 70$~GeV. Thus, the small-$m_H$ expansion
covers the major part of the phase space.
Let us finally mention that for $F_a^{(1),C_A}$
it would be possible to use the ``$m_H\to0$''
even for $q_s^2=-100$~GeV$^2$. In that case
the $q_s$ expansion is only necessary for 
 $p_T=10$~GeV or smaller and $\sqrt{s}$ above $10\,000$~GeV.



\section{\label{sec::results}NNLO one-particle reducible results for
  $gg\to HH$}

In this section we discuss results for the form factors $F^{(1)}_{\rm
  dt1}$, $F^{(1)}_{\rm dt2}$, $F^{(2)}_{\rm dt1}$ and $F^{(2)}_{\rm
  dt2}$, as introduced after Eq.~(\ref{eq::F}).  Sample Feynman
diagrams are shown in Fig.~\ref{fig::diags}.  They can easily be
constructed from the general structure of the
$gg\to H$ vertex from Eq.~(\ref{eq:gghlorentz}), using the
off-shell gluon momentum to connect the two vertices via a gluon
propagator. There are diagrams with either $q_s^2=t$ or $q_s^2=u$.
We additionally must include the one-loop correction to the gluon
propagator, multiplied by a pair of one-loop $gg\to H$ vertices.
This leads to
the following compact formulae for the NNLO form factors,
\begin{equation}
    F^{(2)}_{\rm dt1} = \Tilde{F}^{(2)}_{\rm dt1} (t) +
    \Tilde{F}^{(2)}_{\rm dt1} (u), \quad\quad F^{(2)}_{\rm dt2} =
    \Tilde{F}^{(2)}_{\rm dt2} (t) + \Tilde{F}^{(2)}_{\rm dt2} (u)\,,
\end{equation}
where
\begin{eqnarray}
    \Tilde{F}^{(2)}_{\rm dt1} (t) &=&  F_a^{(0)}(t) \left[ 
      F_a^{(1)}(t) + \frac{1}{2} ~F_a^{(0)}(t) ~\Pi_{gg} (t)
      \right.\nonumber\\ &&\left. \mbox{}
      - \frac{s ~(\epsilon ~(m_H^2-2 p_T^2+t) + 2 p_T^2 ) }{(1-
        2\epsilon)(m_H^2 - s) t} ~F_d^{(1)} (t)  \right]
    \,,
    \label{eq::Ftil1}
\end{eqnarray}
and
\begin{eqnarray}
    \Tilde{F}^{(2)}_{\rm dt2} (t) &=&  F_a^{(0)}(t) \left[ \frac{ p_T^2 }{  t }~
      \left(F_a^{(1)}(t) + \frac{1}{2} ~F_a^{(0)}(t) ~\Pi_{gg} (t)
      \right)
      \right.\nonumber\\ &&\left. \mbox{}
      - \frac{s ~(\epsilon ~(2 p_T^2 -m_H^2 -t) + m_H^2 + t)}{(1-
        2\epsilon)(m_H^2 - s) t} ~F_d^{(1)} (t)  \right]  
    \,.
    \label{eq::Ftil2}
\end{eqnarray}
Here we have used Eq.~\eqref{eq:facdwi} and the fact that $F_d^{(0)}$
is zero. $\Pi_{gg}(q^2)$ is the
transverse part of the one-loop gluon two-point function. Since the
one-loop $gg\to H$ vertex is finite we need $\Pi_{gg}(q^2)$ only up to
its finite part in $\epsilon$. 
From Eqs.~(\ref{eq::Ftil1}) and~(\ref{eq::Ftil2}) it is straightforward
to obtain the corresponding formulae for the NLO (two-loop)
corrections: in the square brackets one has to set $\Pi_{gg}$ and
$F_d^{(1)}$ to zero and replace $F_a^{(1)}$ by $F_a^{(0)}$.

Our result depends on the QCD gauge parameter which is introduced
in the gluon propagator according to
\begin{eqnarray}
  D_g(q) &=& \frac{1}{i} \left(\frac{-g^{\mu\nu} + \xi \frac{q^\mu q^\nu}{-q^2}}{-q^2}\right)
  \,.
\end{eqnarray}

There are several available cross-checks for our calculation.
First we find agreement
with the exact two-loop result obtained in Ref.~\cite{Degrassi:2016vss}.
Furthermore, we perform analytic and
numerical comparisons in the large-$m_t$ limit and cross check
the results for
the form factors $F_{\rm dt1}$ and $F_{\rm dt2}$ obtained in Ref.~\cite{Davies:2019djw}, where an expansion up to order $1/m_t^{8}$ was
performed. In this paper we have obtained expansions of the master integrals  up to order
$1/m_t^{100}$ in the expansion for $m_H\to0$, using their differential
equations.

\begin{figure}[tb]
  \centering
    \begin{tabular}{cc}
    \includegraphics[width=.45\linewidth]{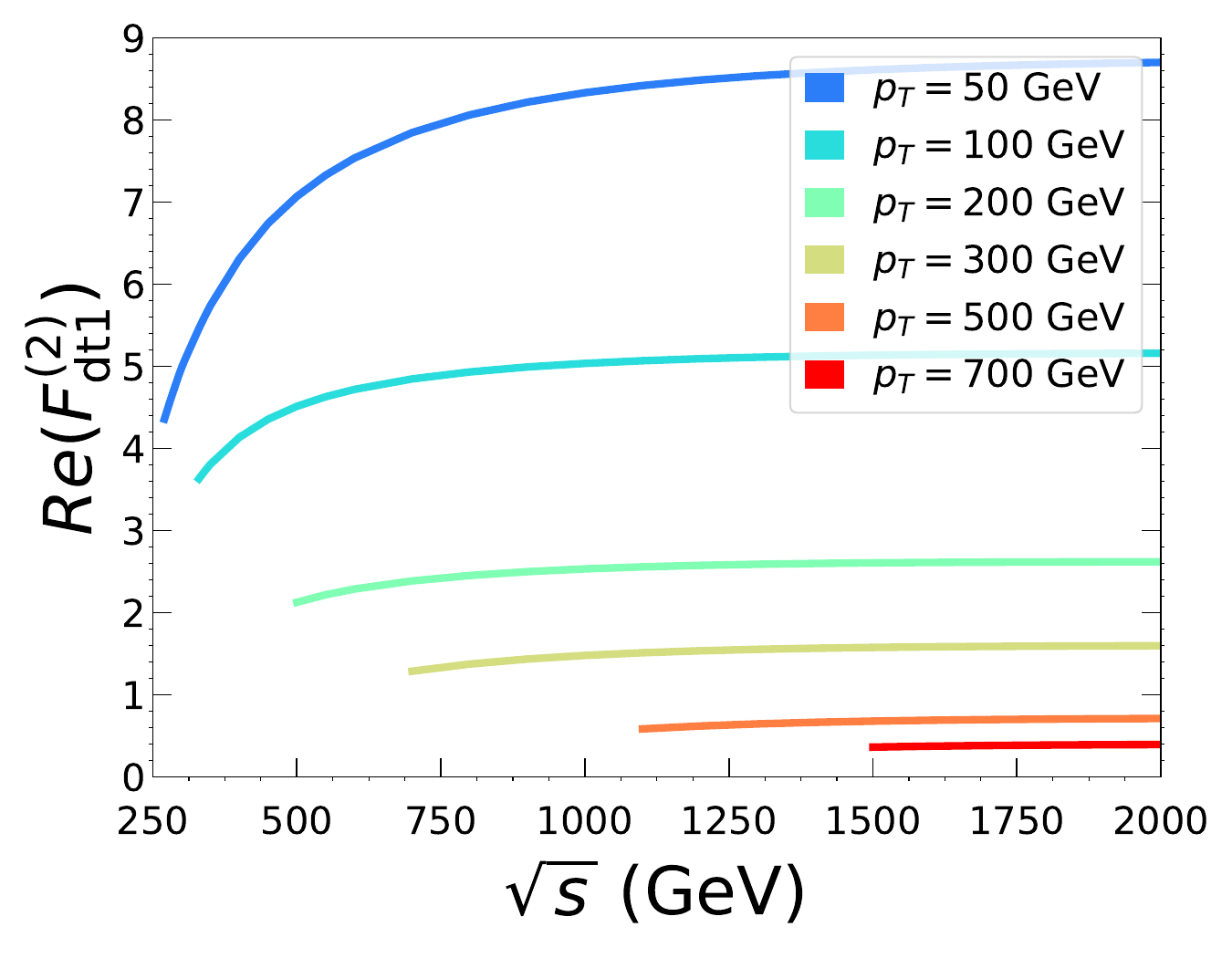} & 
    \includegraphics[width=.45\linewidth]{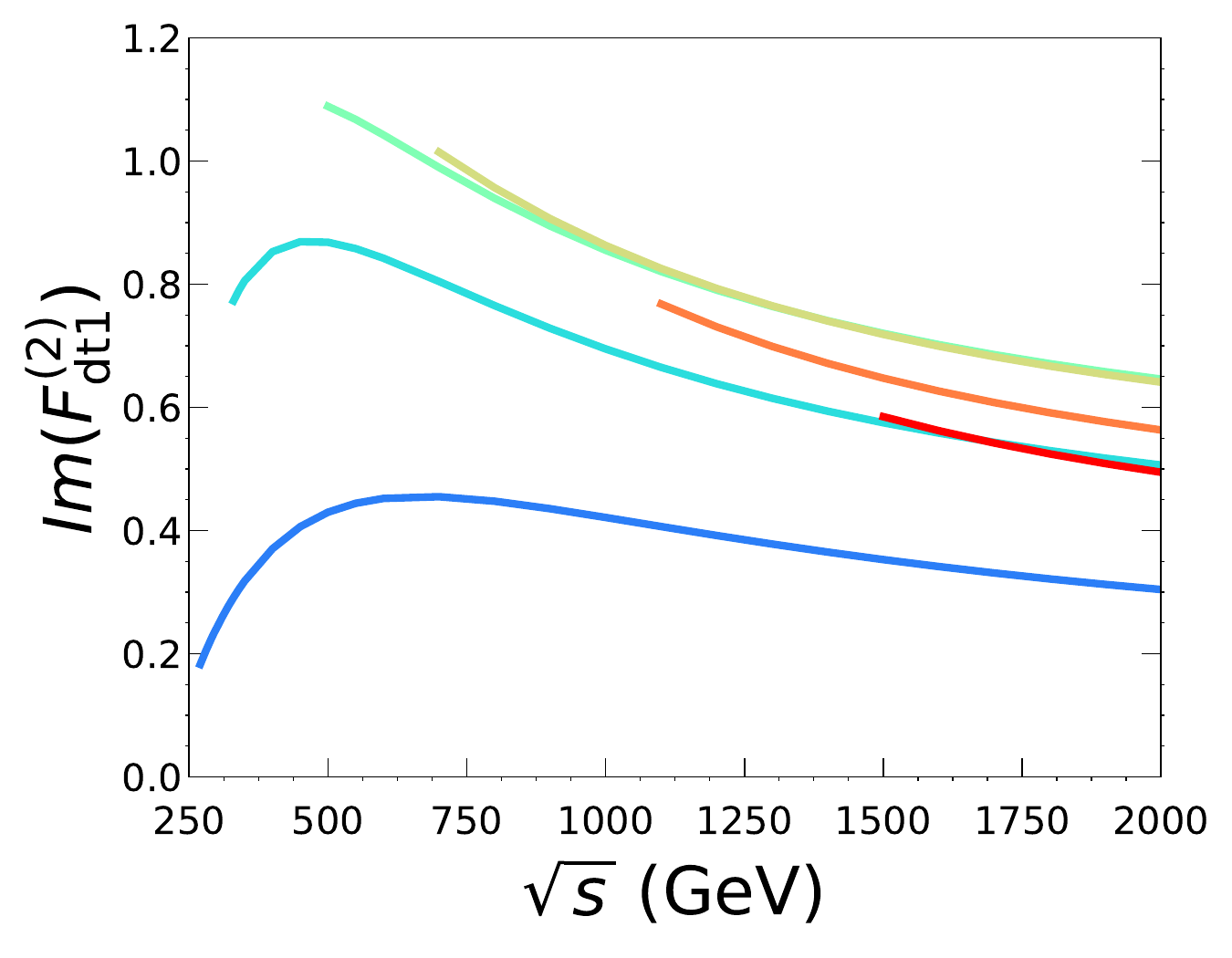}
    \\
    \includegraphics[width=.45\linewidth]{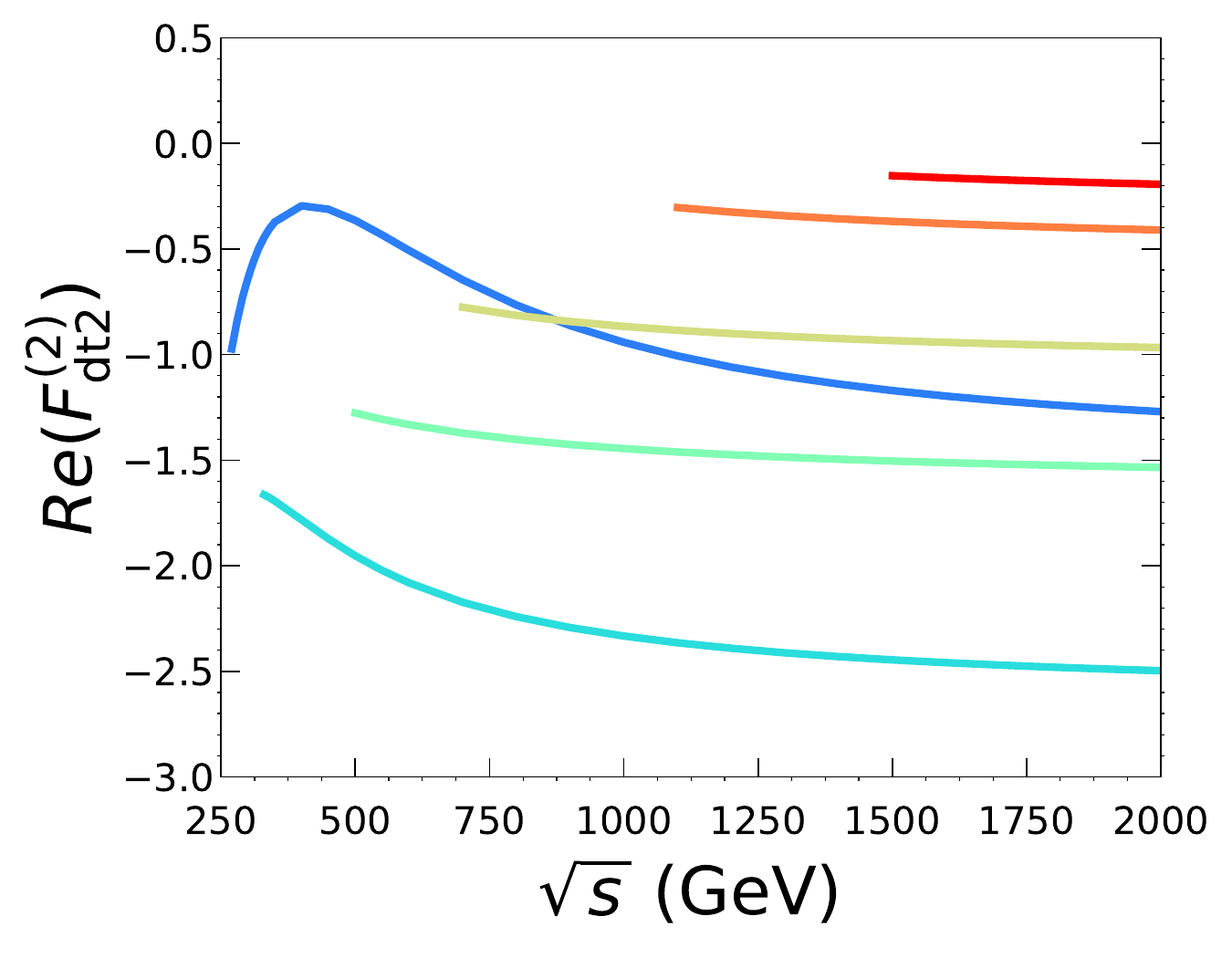} &
    \includegraphics[width=.45\linewidth]{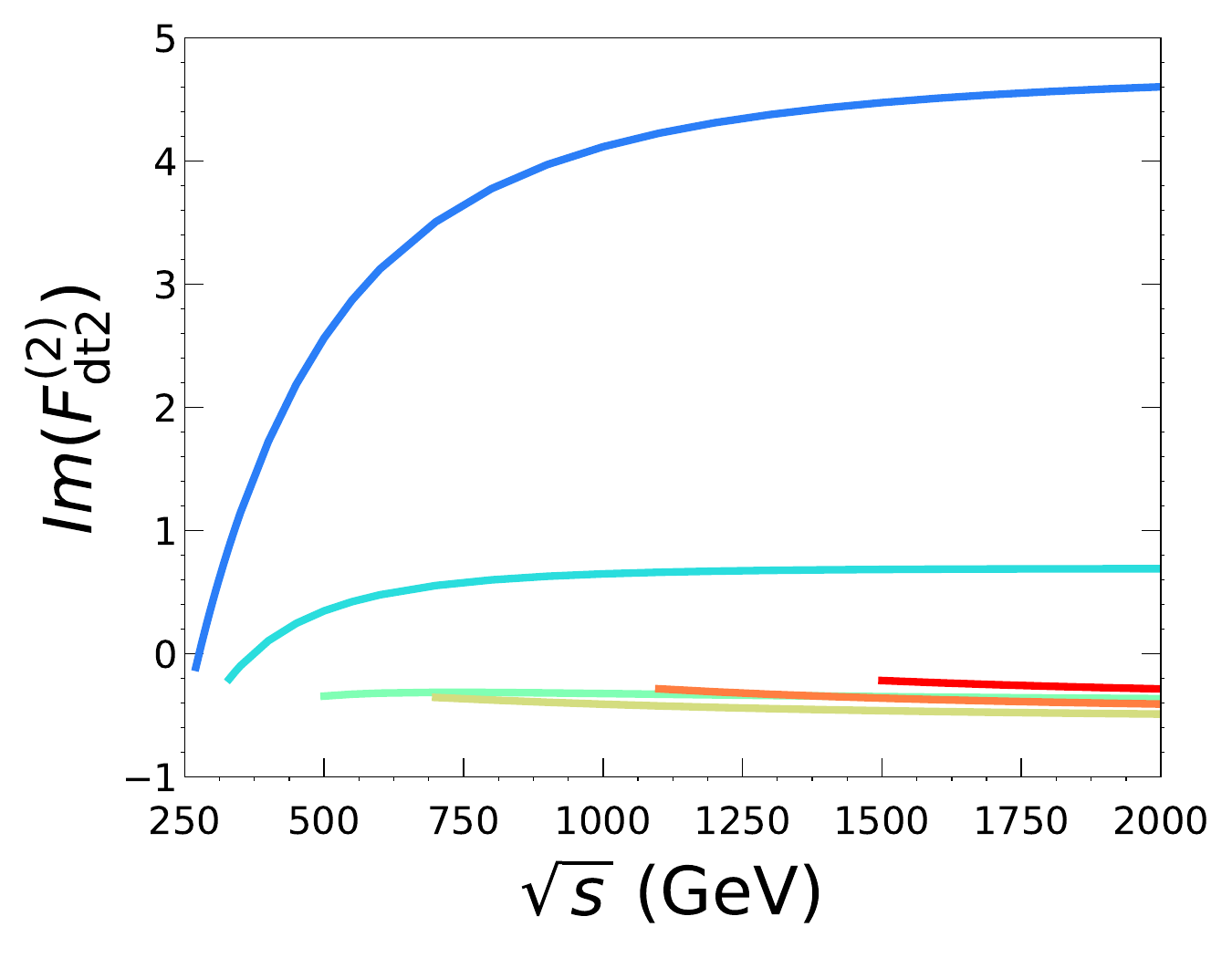}
  \end{tabular}
  \caption{The finite parts of $F^{(2)}_{\rm dt1}$ and $F^{(2)}_{\rm dt2}$ as
    a function of $\sqrt{s}$, for various values of $p_T$. For the
    renormalization scale $\mu^2=m_t^2$ has been chosen, and the gauge parameter
    $\xi = 0$. }\label{fig::F1F2}
\end{figure}

In Fig.~\ref{fig::F1F2} we show the ${\cal O}(\epsilon^0)$ terms of the form
factors $F_{\rm dt1}^{(2)}$ and $F_{\rm dt2}^{(2)}$ as a function of the
partonic centre-of-mass energy $\sqrt{s}$,
for several fixed values of the transverse momentum $p_T$
between $50$~GeV and $700$~GeV.
We note that in order to obtain physical quantities one has to combine the
results presented in this paper with the one-particle irreducible contributions
(which are a work-in-progress),
perform the renormalization and add the contribution from additional
real radiation.
We have implemented our results in a flexible {\tt Mathematica} code 
which can be used for such a combination, where the evaluation of the
one-particle reducible form
factors takes about one second per phase-space point.
In the future we will also provide a high-performance {\tt C++}
implementation of both the one-particle reducible and irreducible NNLO
form factors.

In the ancillary files of this paper we provide all results
which are available analytically. This includes the
$m_H$-expanded one-loop form factor $F_a^{(0)}$
including $m_H^4$ terms,
and the small-$q_s^2$ expansion of the two-loop
form factors $F_a^{(1)}$ and $F_d^{(1)}$ including $s^5$ terms.\footnote{We have computed the expansions up to order $s^{19}$; they can be obtained from the authors upon request.}
The computer-readable code can be downloaded from Ref.~\cite{progdata}.


\section{\label{sec::concl}Conclusions}

In this paper we provide an important ingredient which contributes to the
NNLO virtual
corrections for the process $gg\to HH$, namely the contribution from
one-particle reducible diagrams.  We describe in detail the
calculation of the two-loop $gg\to H$ vertex with an off-shell
gluon, which is used as building block. We obtain analytic results for the expansion around small gluon virtuality and ``semi-analytic'' results for the expansion for small $m_H$. The combination
of both expansions provides a precise approximation of the off-shell $gg\to H$ vertex. Furthermore, we compute the
one-loop corrections including terms of ${\cal O}(\epsilon^2)$.  We
present results for the bare form factors $F_{\rm dt1}$ and $F_{\rm dt2}$ which can
easily be combined with other NNLO ingredients, once they are available.



\section*{Acknowledgements}  

This research was supported by the Deutsche
Forschungsgemeinschaft (DFG, German Research Foundation) under grant 396021762
--- TRR 257 ``Particle Physics Phenomenology after the Higgs Discovery''
and has received funding from the European Research Council (ERC) under
the European Union's Horizon 2020 research and innovation programme grant
agreement 101019620 (ERC Advanced Grant TOPUP).
The work of JD is supported by the STFC Consolidated Grant ST/X000699/1.


%
%



\bibliographystyle{jhep} 
\bibliography{inspire.bib}

\providecommand{\href}[2]{#2}\begingroup\raggedright\begin{thebibliography}{10}

\bibitem{Glover:1987nx}
E.~W.~N. Glover and J.~J. van~der Bij, \emph{{Higgs boson pair production via
  gluon fusion}},
  \href{https://doi.org/10.1016/0550-3213(88)90083-1}{\emph{Nucl. Phys. B}
  {\bfseries 309} (1988) 282}.

\bibitem{Plehn:1996wb}
T.~Plehn, M.~Spira and P.~M. Zerwas, \emph{{Pair production of neutral Higgs
  particles in gluon-gluon collisions}},
  \href{https://doi.org/10.1016/0550-3213(96)00418-X}{\emph{Nucl. Phys. B}
  {\bfseries 479} (1996) 46}
  [\href{https://arxiv.org/abs/hep-ph/9603205}{{\ttfamily hep-ph/9603205}}].

\bibitem{Borowka:2016ehy}
S.~Borowka, N.~Greiner, G.~Heinrich, S.~P. Jones, M.~Kerner, J.~Schlenk et~al.,
  \emph{{Higgs Boson Pair Production in Gluon Fusion at Next-to-Leading Order
  with Full Top-Quark Mass Dependence}},
  \href{https://doi.org/10.1103/PhysRevLett.117.079901}{\emph{Phys. Rev. Lett.}
  {\bfseries 117} (2016) 012001}
  [\href{https://arxiv.org/abs/1604.06447}{{\ttfamily 1604.06447}}].

\bibitem{Borowka:2016ypz}
S.~Borowka, N.~Greiner, G.~Heinrich, S.~P. Jones, M.~Kerner, J.~Schlenk et~al.,
  \emph{{Full top quark mass dependence in Higgs boson pair production at
  NLO}}, \href{https://doi.org/10.1007/JHEP10(2016)107}{\emph{JHEP} {\bfseries
  10} (2016) 107} [\href{https://arxiv.org/abs/1608.04798}{{\ttfamily
  1608.04798}}].

\bibitem{Baglio:2018lrj}
J.~Baglio, F.~Campanario, S.~Glaus, M.~M\"uhlleitner, M.~Spira and
  J.~Streicher, \emph{{Gluon fusion into Higgs pairs at NLO QCD and the top
  mass scheme}},
  \href{https://doi.org/10.1140/epjc/s10052-019-6973-3}{\emph{Eur. Phys. J. C}
  {\bfseries 79} (2019) 459}
  [\href{https://arxiv.org/abs/1811.05692}{{\ttfamily 1811.05692}}].

\bibitem{Bellafronte:2022jmo}
L.~Bellafronte, G.~Degrassi, P.~P. Giardino, R.~Gr\"ober and M.~Vitti,
  \emph{{Gluon fusion production at NLO: merging the transverse momentum and
  the high-energy expansions}},
  \href{https://doi.org/10.1007/JHEP07(2022)069}{\emph{JHEP} {\bfseries 07}
  (2022) 069} [\href{https://arxiv.org/abs/2202.12157}{{\ttfamily
  2202.12157}}].

\bibitem{Davies:2023vmj}
J.~Davies, G.~Mishima, K.~Sch\"onwald and M.~Steinhauser, \emph{{Analytic
  approximations of 2 \textrightarrow{} 2 processes with massive internal
  particles}}, \href{https://doi.org/10.1007/JHEP06(2023)063}{\emph{JHEP}
  {\bfseries 06} (2023) 063}
  [\href{https://arxiv.org/abs/2302.01356}{{\ttfamily 2302.01356}}].

\bibitem{Grigo:2013rya}
J.~Grigo, J.~Hoff, K.~Melnikov and M.~Steinhauser, \emph{{On the Higgs boson
  pair production at the LHC}},
  \href{https://doi.org/10.1016/j.nuclphysb.2013.06.024}{\emph{Nucl. Phys. B}
  {\bfseries 875} (2013) 1} [\href{https://arxiv.org/abs/1305.7340}{{\ttfamily
  1305.7340}}].

\bibitem{Degrassi:2016vss}
G.~Degrassi, P.~P. Giardino and R.~Gr\"ober, \emph{{On the two-loop virtual QCD
  corrections to Higgs boson pair production in the Standard Model}},
  \href{https://doi.org/10.1140/epjc/s10052-016-4256-9}{\emph{Eur. Phys. J. C}
  {\bfseries 76} (2016) 411}
  [\href{https://arxiv.org/abs/1603.00385}{{\ttfamily 1603.00385}}].

\bibitem{Davies:2018ood}
J.~Davies, G.~Mishima, M.~Steinhauser and D.~Wellmann, \emph{{Double-Higgs
  boson production in the high-energy limit: planar master integrals}},
  \href{https://doi.org/10.1007/JHEP03(2018)048}{\emph{JHEP} {\bfseries 03}
  (2018) 048} [\href{https://arxiv.org/abs/1801.09696}{{\ttfamily
  1801.09696}}].

\bibitem{Davies:2018qvx}
J.~Davies, G.~Mishima, M.~Steinhauser and D.~Wellmann, \emph{{Double Higgs
  boson production at NLO in the high-energy limit: complete analytic
  results}}, \href{https://doi.org/10.1007/JHEP01(2019)176}{\emph{JHEP}
  {\bfseries 01} (2019) 176}
  [\href{https://arxiv.org/abs/1811.05489}{{\ttfamily 1811.05489}}].

\bibitem{Bonciani:2018omm}
R.~Bonciani, G.~Degrassi, P.~P. Giardino and R.~Gr\"ober, \emph{{Analytical
  Method for Next-to-Leading-Order QCD Corrections to Double-Higgs
  Production}},
  \href{https://doi.org/10.1103/PhysRevLett.121.162003}{\emph{Phys. Rev. Lett.}
  {\bfseries 121} (2018) 162003}
  [\href{https://arxiv.org/abs/1806.11564}{{\ttfamily 1806.11564}}].

\bibitem{Grober:2017uho}
R.~Gr\"ober, A.~Maier and T.~Rauh, \emph{{Reconstruction of top-quark mass
  effects in Higgs pair production and other gluon-fusion processes}},
  \href{https://doi.org/10.1007/JHEP03(2018)020}{\emph{JHEP} {\bfseries 03}
  (2018) 020} [\href{https://arxiv.org/abs/1709.07799}{{\ttfamily
  1709.07799}}].

\bibitem{Xu:2018eos}
X.~Xu and L.~L. Yang, \emph{{Towards a new approximation for pair-production
  and associated-production of the Higgs boson}},
  \href{https://doi.org/10.1007/JHEP01(2019)211}{\emph{JHEP} {\bfseries 01}
  (2019) 211} [\href{https://arxiv.org/abs/1810.12002}{{\ttfamily
  1810.12002}}].

\bibitem{Wang:2020nnr}
G.~Wang, Y.~Wang, X.~Xu, Y.~Xu and L.~L. Yang, \emph{{Efficient computation of
  two-loop amplitudes for Higgs boson pair production}},
  \href{https://doi.org/10.1103/PhysRevD.104.L051901}{\emph{Phys. Rev. D}
  {\bfseries 104} (2021) L051901}
  [\href{https://arxiv.org/abs/2010.15649}{{\ttfamily 2010.15649}}].

\bibitem{Baglio:2020wgt}
J.~Baglio, F.~Campanario, S.~Glaus, M.~M\"uhlleitner, J.~Ronca and M.~Spira,
  \emph{{$gg\to HH$ : Combined uncertainties}},
  \href{https://doi.org/10.1103/PhysRevD.103.056002}{\emph{Phys. Rev. D}
  {\bfseries 103} (2021) 056002}
  [\href{https://arxiv.org/abs/2008.11626}{{\ttfamily 2008.11626}}].

\bibitem{Bagnaschi:2023rbx}
E.~Bagnaschi, G.~Degrassi and R.~Gr\"ober, \emph{{Higgs boson pair production
  at NLO in the POWHEG approach and the top quark mass uncertainties}},
  \href{https://doi.org/10.1140/epjc/s10052-023-12238-8}{\emph{Eur. Phys. J. C}
  {\bfseries 83} (2023) 1054}
  [\href{https://arxiv.org/abs/2309.10525}{{\ttfamily 2309.10525}}].

\bibitem{Davies:2019djw}
J.~Davies and M.~Steinhauser, \emph{{Three-loop form factors for Higgs boson
  pair production in the large top mass limit}},
  \href{https://doi.org/10.1007/JHEP10(2019)166}{\emph{JHEP} {\bfseries 10}
  (2019) 166} [\href{https://arxiv.org/abs/1909.01361}{{\ttfamily
  1909.01361}}].

\bibitem{Davies:2021kex}
J.~Davies, F.~Herren, G.~Mishima and M.~Steinhauser, \emph{{Real corrections to
  Higgs boson pair production at NNLO in the large top quark mass limit}},
  \href{https://doi.org/10.1007/JHEP01(2022)049}{\emph{JHEP} {\bfseries 01}
  (2022) 049} [\href{https://arxiv.org/abs/2110.03697}{{\ttfamily
  2110.03697}}].

\bibitem{Grigo:2015dia}
J.~Grigo, J.~Hoff and M.~Steinhauser, \emph{{Higgs boson pair production: top
  quark mass effects at NLO and NNLO}},
  \href{https://doi.org/10.1016/j.nuclphysb.2015.09.012}{\emph{Nucl. Phys. B}
  {\bfseries 900} (2015) 412}
  [\href{https://arxiv.org/abs/1508.00909}{{\ttfamily 1508.00909}}].

\bibitem{Davies:2023obx}
J.~Davies, K.~Sch\"onwald and M.~Steinhauser, \emph{{Towards~$gg\to HH$ at
  next-to-next-to-leading order: Light-fermionic three-loop corrections}},
  \href{https://doi.org/10.1016/j.physletb.2023.138146}{\emph{Phys. Lett. B}
  {\bfseries 845} (2023) 138146}
  [\href{https://arxiv.org/abs/2307.04796}{{\ttfamily 2307.04796}}].

\bibitem{Catani:1998bh}
S.~Catani, \emph{{The Singular behavior of QCD amplitudes at two loop order}},
  \href{https://doi.org/10.1016/S0370-2693(98)00332-3}{\emph{Phys. Lett. B}
  {\bfseries 427} (1998) 161}
  [\href{https://arxiv.org/abs/hep-ph/9802439}{{\ttfamily hep-ph/9802439}}].

\bibitem{Nogueira:1991ex}
P.~Nogueira, \emph{{Automatic Feynman Graph Generation}},
  \href{https://doi.org/10.1006/jcph.1993.1074}{\emph{J. Comput. Phys.}
  {\bfseries 105} (1993) 279}.

\bibitem{Gerlach:2022qnc}
M.~Gerlach, F.~Herren and M.~Lang, \emph{{tapir: A tool for topologies,
  amplitudes, partial fraction decomposition and input for reductions}},
  \href{https://doi.org/10.1016/j.cpc.2022.108544}{\emph{Comput. Phys. Commun.}
  {\bfseries 282} (2023) 108544}
  [\href{https://arxiv.org/abs/2201.05618}{{\ttfamily 2201.05618}}].

\bibitem{Harlander:1998cmq}
R.~Harlander, T.~Seidensticker and M.~Steinhauser, \emph{{Complete corrections
  of Order alpha alpha-s to the decay of the Z boson into bottom quarks}},
  \href{https://doi.org/10.1016/S0370-2693(98)00220-2}{\emph{Phys. Lett. B}
  {\bfseries 426} (1998) 125}
  [\href{https://arxiv.org/abs/hep-ph/9712228}{{\ttfamily hep-ph/9712228}}].

\bibitem{Seidensticker:1999bb}
T.~Seidensticker, \emph{{Automatic application of successive asymptotic
  expansions of Feynman diagrams}},  in \emph{{6th International Workshop on
  New Computing Techniques in Physics Research: Software Engineering,
  Artificial Intelligence Neural Nets, Genetic Algorithms, Symbolic Algebra,
  Automatic Calculation}}, 5, 1999,
  \href{https://arxiv.org/abs/hep-ph/9905298}{{\ttfamily hep-ph/9905298}}.

\bibitem{Ruijl:2017dtg}
B.~Ruijl, T.~Ueda and J.~Vermaseren, \emph{{FORM version 4.2}},
  \href{https://arxiv.org/abs/1707.06453}{{\ttfamily 1707.06453}}.

\bibitem{Klappert:2020nbg}
J.~Klappert, F.~Lange, P.~Maierh\"ofer and J.~Usovitsch, \emph{{Integral
  reduction with Kira 2.0 and finite field methods}},
  \href{https://doi.org/10.1016/j.cpc.2021.108024}{\emph{Comput. Phys. Commun.}
  {\bfseries 266} (2021) 108024}
  [\href{https://arxiv.org/abs/2008.06494}{{\ttfamily 2008.06494}}].

\bibitem{Lee:2013mka}
R.~N. Lee, \emph{{LiteRed 1.4: a powerful tool for reduction of multiloop
  integrals}}, \href{https://doi.org/10.1088/1742-6596/523/1/012059}{\emph{J.
  Phys. Conf. Ser.} {\bfseries 523} (2014) 012059}
  [\href{https://arxiv.org/abs/1310.1145}{{\ttfamily 1310.1145}}].

\bibitem{Remiddi:1999ew}
E.~Remiddi and J.~A.~M. Vermaseren, \emph{{Harmonic polylogarithms}},
  \href{https://doi.org/10.1142/S0217751X00000367}{\emph{Int. J. Mod. Phys. A}
  {\bfseries 15} (2000) 725}
  [\href{https://arxiv.org/abs/hep-ph/9905237}{{\ttfamily hep-ph/9905237}}].

\bibitem{Fael:2021kyg}
M.~Fael, F.~Lange, K.~Sch\"onwald and M.~Steinhauser, \emph{{A semi-analytic
  method to compute Feynman integrals applied to four-loop corrections to the $
  \overline{\mathrm{MS}} $-pole quark mass relation}},
  \href{https://doi.org/10.1007/JHEP09(2021)152}{\emph{JHEP} {\bfseries 09}
  (2021) 152} [\href{https://arxiv.org/abs/2106.05296}{{\ttfamily
  2106.05296}}].

\bibitem{Fael:2022rgm}
M.~Fael, F.~Lange, K.~Sch\"onwald and M.~Steinhauser, \emph{{Massive Vector
  Form Factors to Three Loops}},
  \href{https://doi.org/10.1103/PhysRevLett.128.172003}{\emph{Phys. Rev. Lett.}
  {\bfseries 128} (2022) 172003}
  [\href{https://arxiv.org/abs/2202.05276}{{\ttfamily 2202.05276}}].

\bibitem{Fael:2022miw}
M.~Fael, F.~Lange, K.~Sch\"onwald and M.~Steinhauser, \emph{{Singlet and
  nonsinglet three-loop massive form factors}},
  \href{https://doi.org/10.1103/PhysRevD.106.034029}{\emph{Phys. Rev. D}
  {\bfseries 106} (2022) 034029}
  [\href{https://arxiv.org/abs/2207.00027}{{\ttfamily 2207.00027}}].

\bibitem{Fael:2023zqr}
M.~Fael, F.~Lange, K.~Sch\"onwald and M.~Steinhauser, \emph{{Massive three-loop
  form factors: Anomaly contribution}},
  \href{https://doi.org/10.1103/PhysRevD.107.094017}{\emph{Phys. Rev. D}
  {\bfseries 107} (2023) 094017}
  [\href{https://arxiv.org/abs/2302.00693}{{\ttfamily 2302.00693}}].

\bibitem{Spira:1995rr}
M.~Spira, A.~Djouadi, D.~Graudenz and P.~M. Zerwas, \emph{{Higgs boson
  production at the LHC}},
  \href{https://doi.org/10.1016/0550-3213(95)00379-7}{\emph{Nucl. Phys. B}
  {\bfseries 453} (1995) 17}
  [\href{https://arxiv.org/abs/hep-ph/9504378}{{\ttfamily hep-ph/9504378}}].

\bibitem{Harlander:2005rq}
R.~Harlander and P.~Kant, \emph{{Higgs production and decay: Analytic results
  at next-to-leading order QCD}},
  \href{https://doi.org/10.1088/1126-6708/2005/12/015}{\emph{JHEP} {\bfseries
  12} (2005) 015} [\href{https://arxiv.org/abs/hep-ph/0509189}{{\ttfamily
  hep-ph/0509189}}].

\bibitem{Anastasiou:2006hc}
C.~Anastasiou, S.~Beerli, S.~Bucherer, A.~Daleo and Z.~Kunszt, \emph{{Two-loop
  amplitudes and master integrals for the production of a Higgs boson via a
  massive quark and a scalar-quark loop}},
  \href{https://doi.org/10.1088/1126-6708/2007/01/082}{\emph{JHEP} {\bfseries
  01} (2007) 082} [\href{https://arxiv.org/abs/hep-ph/0611236}{{\ttfamily
  hep-ph/0611236}}].

\bibitem{Aglietti:2006tp}
U.~Aglietti, R.~Bonciani, G.~Degrassi and A.~Vicini, \emph{{Analytic Results
  for Virtual QCD Corrections to Higgs Production and Decay}},
  \href{https://doi.org/10.1088/1126-6708/2007/01/021}{\emph{JHEP} {\bfseries
  01} (2007) 021} [\href{https://arxiv.org/abs/hep-ph/0611266}{{\ttfamily
  hep-ph/0611266}}].

\bibitem{Harlander:2009bw}
R.~V. Harlander and K.~J. Ozeren, \emph{{Top mass effects in Higgs production
  at next-to-next-to-leading order QCD: Virtual corrections}},
  \href{https://doi.org/10.1016/j.physletb.2009.08.012}{\emph{Phys. Lett. B}
  {\bfseries 679} (2009) 467}
  [\href{https://arxiv.org/abs/0907.2997}{{\ttfamily 0907.2997}}].

\bibitem{Ablinger:2018zwz}
J.~Ablinger, J.~Bl\"umlein, P.~Marquard, N.~Rana and C.~Schneider,
  \emph{{Automated Solution of First Order Factorizable Systems of Differential
  Equations in One Variable}},
  \href{https://doi.org/10.1016/j.nuclphysb.2018.12.010}{\emph{Nucl. Phys. B}
  {\bfseries 939} (2019) 253}
  [\href{https://arxiv.org/abs/1810.12261}{{\ttfamily 1810.12261}}].

\bibitem{Blumlein:1998if}
{Bl\"umlein, Johannes and Kurth, Stefan}, \emph{{Harmonic sums and Mellin
  transforms up to two loop order}},
  \href{https://doi.org/10.1103/PhysRevD.60.014018}{\emph{Phys. Rev. D}
  {\bfseries 60} (1999) 014018}
  [\href{https://arxiv.org/abs/hep-ph/9810241}{{\ttfamily hep-ph/9810241}}].

\bibitem{Vermaseren:1998uu}
J.~A.~M. Vermaseren, \emph{{Harmonic sums, Mellin transforms and integrals}},
  \href{https://doi.org/10.1142/S0217751X99001032}{\emph{Int. J. Mod. Phys. A}
  {\bfseries 14} (1999) 2037}
  [\href{https://arxiv.org/abs/hep-ph/9806280}{{\ttfamily hep-ph/9806280}}].

\bibitem{Blumlein:2009ta}
{Bl\"umlein, Johannes}, \emph{{Structural Relations of Harmonic Sums and Mellin
  Transforms up to Weight w = 5}},
  \href{https://doi.org/10.1016/j.cpc.2009.07.004}{\emph{Comput. Phys. Commun.}
  {\bfseries 180} (2009) 2218}
  [\href{https://arxiv.org/abs/0901.3106}{{\ttfamily 0901.3106}}].

\bibitem{Ablinger:2009ovq}
J.~Ablinger, \emph{{A Computer Algebra Toolbox for Harmonic Sums Related to
  Particle Physics}},  Master's thesis, Linz U., 2009.

\bibitem{Ablinger:2011te}
{Ablinger, Jakob and Bl\"umlein, Johannes and Schneider, Carsten},
  \emph{{Harmonic Sums and Polylogarithms Generated by Cyclotomic
  Polynomials}}, \href{https://doi.org/10.1063/1.3629472}{\emph{J. Math. Phys.}
  {\bfseries 52} (2011) 102301}
  [\href{https://arxiv.org/abs/1105.6063}{{\ttfamily 1105.6063}}].

\bibitem{Ablinger:2012ufz}
J.~Ablinger, \emph{{Computer Algebra Algorithms for Special Functions in
  Particle Physics}}, Ph.D. thesis, Linz U., 4, 2012.
\newblock \href{https://arxiv.org/abs/1305.0687}{{\ttfamily 1305.0687}}.

\bibitem{Ablinger:2013eba}
J.~Ablinger, J.~Bl\"umlein and C.~Schneider, \emph{{Generalized Harmonic,
  Cyclotomic, and Binomial Sums, their Polylogarithms and Special Numbers}},
  \href{https://doi.org/10.1088/1742-6596/523/1/012060}{\emph{J. Phys. Conf.
  Ser.} {\bfseries 523} (2014) 012060}
  [\href{https://arxiv.org/abs/1310.5645}{{\ttfamily 1310.5645}}].

\bibitem{Ablinger:2013cf}
J.~Ablinger, J.~Bl\"umlein and C.~Schneider, \emph{{Analytic and Algorithmic
  Aspects of Generalized Harmonic Sums and Polylogarithms}},
  \href{https://doi.org/10.1063/1.4811117}{\emph{J. Math. Phys.} {\bfseries 54}
  (2013) 082301} [\href{https://arxiv.org/abs/1302.0378}{{\ttfamily
  1302.0378}}].

\bibitem{Ablinger:2014bra}
J.~Ablinger, J.~Bl\"umlein, C.~G. Raab and C.~Schneider, \emph{{Iterated
  Binomial Sums and their Associated Iterated Integrals}},
  \href{https://doi.org/10.1063/1.4900836}{\emph{J. Math. Phys.} {\bfseries 55}
  (2014) 112301} [\href{https://arxiv.org/abs/1407.1822}{{\ttfamily
  1407.1822}}].

\bibitem{Ablinger:2014rba}
J.~Ablinger, \emph{{The package HarmonicSums: Computer Algebra and Analytic
  aspects of Nested Sums}},
  \href{https://doi.org/10.22323/1.211.0019}{\emph{PoS} {\bfseries LL2014}
  (2014) 019} [\href{https://arxiv.org/abs/1407.6180}{{\ttfamily 1407.6180}}].

\bibitem{Ablinger:2015gdg}
J.~Ablinger, \emph{{Discovering and Proving Infinite Binomial Sums
  Identities}},
  \href{https://doi.org/10.1080/10586458.2015.1116028}{\emph{Exper. Math.}
  {\bfseries 26} (2016) 62} [\href{https://arxiv.org/abs/1507.01703}{{\ttfamily
  1507.01703}}].

\bibitem{Ablinger:2018cja}
J.~Ablinger, \emph{{Computing the Inverse Mellin Transform of Holonomic
  Sequences using Kovacic\textquoteright{}s Algorithm}},
  \href{https://doi.org/10.22323/1.290.0001}{\emph{PoS} {\bfseries RADCOR2017}
  (2018) 001} [\href{https://arxiv.org/abs/1801.01039}{{\ttfamily
  1801.01039}}].

\bibitem{sigmaI}
C.~Schneider, \emph{{Symbolic summation assists combinatorics}},
  {\emph{Seminaire Lotharingien de Combinatoire} {\bfseries 56} (2007) 1}.

\bibitem{sigmaII}
C.~Schneider, \emph{Term algebras, canonical representations and difference
  ring theory for symbolic summation}, {\emph{CoRR} {\bfseries abs/2102.01471}
  (2021) } [\href{https://arxiv.org/abs/2102.01471}{{\ttfamily 2102.01471}}].

\bibitem{Jantzen:2012mw}
B.~Jantzen, A.~V. Smirnov and V.~A. Smirnov, \emph{{Expansion by regions:
  revealing potential and Glauber regions automatically}},
  \href{https://doi.org/10.1140/epjc/s10052-012-2139-2}{\emph{Eur. Phys. J. C}
  {\bfseries 72} (2012) 2139}
  [\href{https://arxiv.org/abs/1206.0546}{{\ttfamily 1206.0546}}].

\bibitem{RISC4711}
C.~Schneider, \emph{{Simplifying Multiple Sums in Difference Fields}},  in
  \emph{{Computer Algebra in Quantum Field Theory: Integration, Summation and
  Special Functions}}, C.~Schneider and J.~Bluemlein, eds., Texts and
  Monographs in Symbolic Computation, pp.~325--360, Springer, (2013),
  \href{https://www.doi.org/10.1007/978-3-7091-1616-6_14}{https://www.doi.org/10.1007/978-3-7091-1616-6_14}.

\bibitem{RISC4826}
J.~Ablinger, J.~Blümlein, P.~Marquard, N.~Rana and C.~Schneider, \emph{{Three
  loop QCD corrections to heavy quark form factors}},  in \emph{{Proc. ACAT
  2019}}, vol.~1525 of \emph{J. Phys.: Conf. Ser.}, pp.~1--10, 2020,
  \href{https://www.doi.org/10.1088/1742-6596/1525/1/012018}{https://www.doi.org/10.1088/1742-6596/1525/1/012018}.

\bibitem{Liu:2022chg}
X.~Liu and Y.-Q. Ma, \emph{{AMFlow: A Mathematica package for Feynman integrals
  computation via auxiliary mass flow}},
  \href{https://doi.org/10.1016/j.cpc.2022.108565}{\emph{Comput. Phys. Commun.}
  {\bfseries 283} (2023) 108565}
  [\href{https://arxiv.org/abs/2201.11669}{{\ttfamily 2201.11669}}].

\bibitem{progdata}
\verb|https://www.ttp.kit.edu/preprints/2024/ttp24-016/|.

\end{thebibliography}\endgroup

\end{document}